\documentclass[%
 twocolumn,
superscriptaddress,
preprintnumbers,
nofootinbib,
 amsmath,amssymb,
 aps,
prd,
longbibliography,
floatfix,
]{revtex4-2}

\usepackage{graphicx}
\usepackage{dcolumn}
\usepackage{bm}
\usepackage[dvipsnames]{xcolor}
\usepackage[colorlinks]{hyperref}
\usepackage{todonotes}

\hypersetup{linkcolor=BrickRed,citecolor=Green,
filecolor=Mulberry,
urlcolor=NavyBlue,
menucolor=BrickRed,
runcolor=Mulberry
}

\usepackage{tikz}
\usetikzlibrary{tikzmark}
\usetikzlibrary{positioning}
\usepackage[normalem]{ulem} 

\definecolor{lime}{HTML}{A6CE39}
\DeclareRobustCommand{\orcidicon}{\hspace{-1mm}
	\begin{tikzpicture}
	\draw[lime, fill=lime] (0,0) 
	circle [radius=0.16] 
	node[white] {{\fontfamily{qag}\selectfont \tiny \,ID}};
	\draw[white, fill=white] (-0.0525,0.095) 
	circle [radius=0.007];
	\end{tikzpicture}
	\hspace{-3mm}
}

\foreach \x in {A, ..., Z}{\expandafter\xdef\csname orcid\x\endcsname{\noexpand\href{https://orcid.org/\csname orcidauthor\x\endcsname}
			{\noexpand\orcidicon}}
}

\newcommand{\bN}{\overline{N}}
\newcommand{\bF}{\overline{F}}
\newcommand{\F}{\mathcal{F}}
\newcommand{\x}{\mathrm{x}}

\newcommand{\vrho}{\varrho}
\newcommand{\bvrho}{\bar{\varrho}}
\newcommand{\dd}{\mathrm{d}}
\newcommand{\Tr}{\mathrm{Tr}}

\newcommand{\yML}{\mathsf{y}}

\renewcommand{\vec}{\mathbf}

\begin{document}
\preprint{N3AS-24-030}

\title{Asymptotic-state prediction for fast flavor transformation in neutron star mergers}

\author{Sherwood Richers\orcidA{}}
\email{richers@utk.edu}
\affiliation{Department of Physics, University of Tennessee Knoxville, Knoxville, TN 37996, USA}

\author{Julien Froustey\orcidB{}}
\email{jfroustey@berkeley.edu}
\affiliation{Department of Physics, University of California Berkeley, Berkeley, CA 94720, USA}
\affiliation{Department of Physics, North Carolina State University, Raleigh, NC 27695, USA}

\author{Somdutta Ghosh\orcidC{}}
\email{Somdutta.Ghosh@unh.edu}
\affiliation{Department of Physics \& Astronomy, University of New Hampshire, 9 Library Way, Durham, NH 03824, USA}

\author{Francois Foucart\orcidD{}}
\email{Francois.Foucart@unh.edu}
\affiliation{Department of Physics \& Astronomy, University of New Hampshire, 9 Library Way, Durham, NH 03824, USA}

\author{Javier Gomez\orcidE{}}
\affiliation{Department of Physics, University of Tennessee Knoxville, Knoxville, TN 37996, USA}

\date{\today}

\begin{abstract}
Neutrino flavor instabilities appear to be omnipresent in dense astrophysical environments, thus presenting a challenge to large-scale simulations of core-collapse supernovae and neutron star mergers (NSMs). Subgrid models offer a path forward, but require an accurate determination of the local outcome of such conversion phenomena. Focusing on “fast” instabilities, related to the existence of a crossing between neutrino and antineutrino angular distributions, we consider a range of analytical mixing schemes, including a new, fully three-dimensional one, and also introduce a new machine learning (ML) model. We compare the accuracy of these models with the results of several thousands of local dynamical calculations of neutrino evolution from the conditions extracted from classical NSM simulations. Our ML model shows good overall performance, but struggles to generalize to conditions from a NSM simulation not used for training. The multidimensional analytic model performs and generalizes even better, while other analytic models (which assume axisymmetric neutrino distributions) do not have reliably high performances, as they notably fail as expected to account for effects resulting from strong anisotropies. The ML and analytic subgrid models extensively tested here are both promising, with different computational requirements and sources of systematic errors.
\end{abstract}

\maketitle


\section{Introduction}

Neutrino flavor transformation, an intrinsically quantum phenomenon first discovered with atmospheric~\cite{Kajita:2016cak} and solar~\cite{McDonald:2016ixn} neutrinos, is now well-established. The mixing parameters in the three-flavor scenario have been measured with increasing accuracy in the last decades~\cite{PDG_2024}, or are within reach of existing and future neutrino experiments (CP-phase, $\theta_\mathrm{23}$ octant, mass ordering). In dense astrophysical environments like core-collapse supernovae (CCSNe) or neutron star mergers (NSMs), the very high matter and neutrino densities source effective potentials which necessitate a quantum description of neutrino kinetics \cite{sigl_GeneralKineticDescription_1993,volpe2015neutrino,vlasenko2014neutrino,blaschke_NeutrinoQuantumKinetic_2016}. This wrinkle results in a rich and complex phenomenology of neutrino flavor conversion (see~\cite{Tamborra_NewDevelopmentsFlavor_2021,Capozzi_NeutrinoFlavorConversions_2022,Volpe_review_2023} for recent reviews).

In particular, “fast” flavor instabilities (FFIs), which are associated to the existence of an (electron $-$ heavy-leptonic flavor) neutrino lepton number (ELN$-$XLN) angular crossing, have been shown to be very common in dense environments (see e.g.,~\cite{Wu_FastNeutrinoConversions_2017,Abbar_FastNeutrinoFlavor_2018,Morinaga_FastNeutrinoflavorConversion_2020,M.d.azari_LinearAnalysisFastpairwise_2019,Nagakura_WhereWhenWhy_2021,Richers_evaluating_2022,richers_FastFlavorTransformations_2022,fiorillo2024theory}). The neutrino distributions in existing NSM simulations imply unstable modes that grow on timescales of nanoseconds and lengthscales of millimeters. Multidimensional direct simulation of neutrino quantum kinetics on large scales is computationally challenging 
\cite{nagakura_TimeDependentQuasisteadyFeatures_2022,Nagakura:2022xwe,Nagakura:2023wbf} even if such strong instabilities are not fully realized (e.g., \cite{Fiorillo:2024qbl}).

One path forward consists in designing an appropriate subgrid model, which can be incorporated with limited cost in hydrodynamic codes (see, for instance, \cite{li_NeutrinoFastFlavor_2021,just_FastNeutrinoConversion_2022,fernandez_fast_2022,Ehring_FastNeutrinoFlavor_2023,Ehring_HelpHinderSupernova_2023,Nagakura_BGK}). These models must first identify the locations and growth rates of instabilities, which can be done with linear stability analysis~\cite{Izaguirre_FastPairwiseConversion_2017,Capozzi_Classifying_2017,Morinaga_LinearStabilityAnalysis_2018,Froustey_LSA_2023}. They must then determine the asymptotic state reached after the FFI, for which a general analytic solution is not yet known. 
Local, dynamical simulations of neutrino evolution in NSM-like environments can be used to estimate the asymptotic state on small scales. Simulation codes performing such a task are widely available with varying numerical techniques and approximations (e.g., \cite{richers_code_2022,George:2022lwg,richers_ParticleincellSimulationNeutrino_2021,Kato_2021,Grohs:2023pgq}). Although a number of approximate mixing schemes have been proposed, the most sophisticated of them are restricted to axially symmetric situations (e.g., \cite{just_FastNeutrinoConversion_2022,bhattacharyya_FastFlavorDepolarization_2021,Bhattacharyya_ElaboratingFate_2022,nagakura_TimeDependentQuasisteadyFeatures_2022,Zaizen:2022cik,Zaizen_CharacterizingQuasisteadyStates_2023,Xiong:2023vcm,Xiong:2024tac}).

An attractive alternative to analytic subgrid models is a machine learning (ML) model trained to reproduce the results of local simulations of flavor instability.
ML technology applied to neutrino flavor instabilities has been pioneered by Abbar et al., with models dedicated to the detection of instabilities, trained with artificial data~\cite{Abbar_ML_detect_1}, more physical distributions~\cite{Abbar_ML_detect_2} and extended to non-axisymmetric crossings in~\cite{Abbar_ML_detect_nonaxisym}. The capability of ML models to predict the outcome of FFIs, based on an axisymmetric analytical prescription developed in~\cite{Xiong:2023vcm}, was discussed in~\cite{Abbar_ML_outcome}, with a multi-energy generalization in~\cite{Abbar_ML_outcome_multiE}. These works have demonstrated that ML can in principle perform very well under the constraints imposed, but cannot be directly applied to fully relativistic and multidimensional hydrodynamic simulations without further improvements, notably removing symmetry simplifications and expanding the parameter space of the training set. In addition, it is presently unclear whether ML models in general circumstances actually have superior performance over other analytic subgrid models.

It is also important to note that the majority of multidimensional simulations of CCSNe and NSMs evolve the neutrino radiation field based on angular moments \cite{Thorne:1981nvt,Shibata:2011kx,Cardall:2012at}. While some subgrid models have been proposed to predict the angular structure of the post-instability distribution \cite{Bhattacharyya_ElaboratingFate_2022,Zaizen_CharacterizingQuasisteadyStates_2023,Zaizen:2022cik,Xiong:2023vcm}, others are built to predict post-instability moments using only pre-instability moments as inputs (e.g., \cite{just_FastNeutrinoConversion_2022,Abbar_ML_outcome,Ehring_HelpHinderSupernova_2023}). The latter are most directly applicable to current state-of-the-art global simulations.

There are nevertheless a few caveats to take into account. First, the previously mentioned models all start from a “classical” configuration, i.e., without flavor mixing, which consists in the distributions (or the angular moments) of each flavor of (anti)neutrinos. The model then predicts the final distributions or moments, once again for each flavor. For instance, ($f_{\nu_e},f_{\nu_\mu},f_{\nu_\tau}$) become ($f'_{\nu_e},f'_{\nu_\mu},f'_{\nu_\tau}$). This approach neglects any quantum coherence in both the initial and final states. In the density matrix formalism, this amounts to considering that the final density matrix is essentially flavor-diagonal. There are nevertheless examples of systems with maintain some flavor coherence, for which an “effectively classical” description is not adequate (see e.g.,~\cite{Duan_FlavorIsospinWaves_2021,Fiorillo_FlavorSolitons_2023}). 

Second, the use of \emph{local} asymptotic-state subgrid models must be considered with caution. As discussed in~\cite{Johns:2024dbe}, the underlying idea of such subgrid models in a large-scale simulation is to instantaneously impose the final state that we predict here, whenever the flavor configuration is unstable, using the fact that the timescale of fast flavor conversion $\Delta t_\mathrm{FFI}$ is much smaller than the time steps of a simulation $\Delta t_\mathrm{simu}$. This procedure is then reiterated after a simulation time step, and so on. However, there is an internal inconsistency as the evolution between two simulation time steps is entirely classical, although fast flavor conversions should be, by assumption, potentially taking place during $\Delta t_\mathrm{simu} \gg \Delta t_\mathrm{FFI}$. Since the FFI is expected to erase ELN-XLN crossings, one might expect that in an actual system neutrinos always stay at the edge of instability, a regime where flavor conversion might be different from the cases we consider here, as recently explored in~\cite{Fiorillo:2024qbl}. In addition, gradients allow for an asymptotic state that is qualitatively different from one predicted by simulations with periodic boundary conditions (e.g., \cite{sigl2022simulations,Zaizen_CharacterizingQuasisteadyStates_2023}), even including completely swapping flavors instead of relaxing to a mixed equilibrium under certain boundary conditions \cite{Nagakura:2023wbf,zaizen_fastswap_2024}. Nevertheless, asymptotic-state subgrid models present the tremendous advantage of being relatively easy to implement in existing large-scale simulations. Such models can provide insight into the implications of local instabilities on the evolution of CCSNe and NSMs when a fully consistent treatment of flavor transformation is still not yet achievable.

Third, it is well known that representing neutrino distributions with a small number of angular moments can induce sizeable errors (e.g., \cite{murchikova_AnalyticClosuresM1_2017} for classical transport, and~\cite{Johns:2019izj,Johns:2020qsk} in the quantum case). In the context of NSMs, where strong anisotropies and inhomogeneities can break the assumptions that go into constructing an analytic closure, the maximum entropy closure used in the following cannot represent the true angular structure of the radiation field~\cite{Richers_Rank3_2020,Foucart_Evaluating_2018}. The FFI generally involves strong angular dependence in the amount of flavor transformation, so the post-instability distributions are far from the maximum-entropy distributions assumed here (e.g.,~\cite{richers_ParticleincellSimulationNeutrino_2021,Wu_Collective1DBox_2021,Nagakura_BasicCharacteristics_2023,Xiong:2023vcm}). Just as with classical radiation transport, a fully angle-dependent treatment of the FFI is needed to capture all of the details, though such treatments are naturally much more expensive. That being said, moment methods for quantum kinetics work surprisingly well, both in terms of analytically predicting flavor instabilities \cite{Froustey_LSA_2023} and directly simulating them \cite{Grohs_NeutrinoFastFlavor_2022,Grohs:2023pgq}, especially in the context of recent improvements to analytic treatments of quantum closures \cite{Froustey:2024sgz}, cementing moment methods as an attractive, inexpensive alternative to fully multi-dimensional quantum kinetics.

In this paper, we perform a systematic assessment of the performance of existing mixing schemes in actual NSM environments, based on tens of thousands of local, three-dimensional simulations of FFIs. We go beyond the standard axisymmetric assumption by generalizing mixing prescriptions based on flavor equilibration in a particular angular domain to a generic three-dimensional system, which is a necessary step forward to maintain a reasonable accuracy on the direction of (anti)neutrino fluxes. We also introduce a new ML model, specifically designed for relativistic settings, and evaluate its performance on different NSM simulation snapshots.

The most directly comparable ML model in the literature is that of \cite{Abbar_ML_outcome}, in which the authors develop a neural network to predict the outcome of the FFI assuming axisymmetric neutrino distributions. In this work the authors use a collection of data points both generated analytically via the approximate “Power-1/2” scheme~\cite{Xiong:2023vcm} and from direct simulation, and train a neural network with a single hidden layer to reproduce the post-instability moments with accuracies of a few percent. We build on this work in a few important ways. We consider general anisotropies, allowing distributions from all neutrino flavors to have fluxes in arbitrary directions. We construct our model to take as input only relativistically invariant quantities, produce results that are also invariant under rotations and boosts, and do not assume knowledge of ELN conservation, crossing location, or relative crossing depth. Finally, our training, test, and validation datasets are drawn from a combination of randomly-generated neutrino distributions and distributions extracted from 3D general-relativistic simulations of NSMs.

This paper is organized as follows. In Sec.~\ref{sec:ML_model}, we present our machine learning model and the datasets used for its training and testing. In Sec.~\ref{sec:mixing_schemes}, we review existing mixing schemes previously used in the literature and generalize them to a multidimensional environment, including introducing a straightforward and performant three-dimensional, non-axisymmetric model. Our main results, comparing the performance of all these subgrid models on various datasets, are gathered in Sec.~\ref{sec:performance}. We summarize and conclude in Sec.~\ref{sec:conclusion}.

\section{Machine Learning Model}
\label{sec:ML_model}

\subsection{Design}

\begin{figure*}
    \centering
    \includegraphics[width=\linewidth]{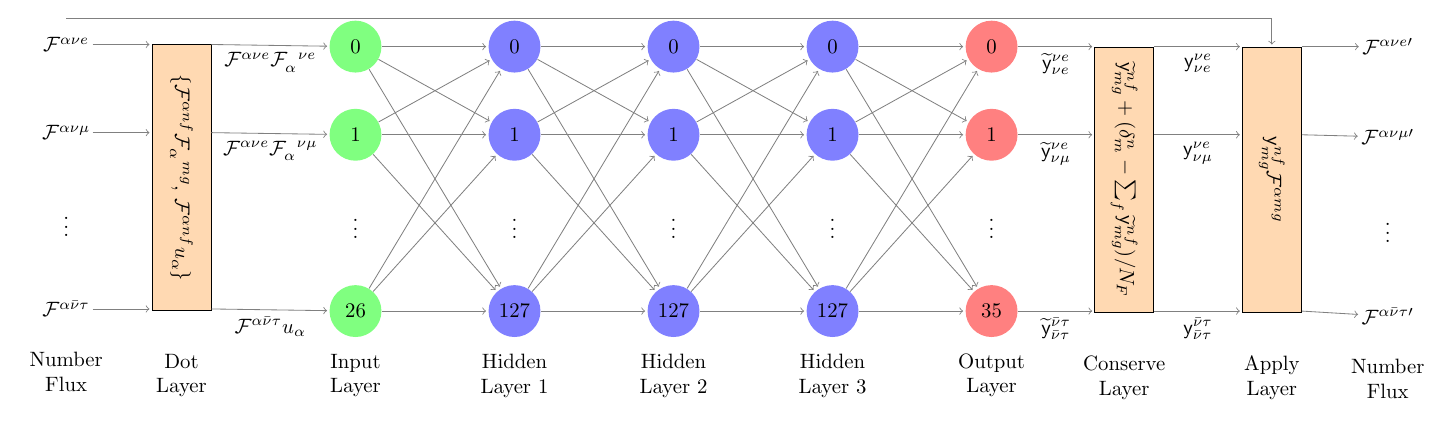}
    \caption{Neural network architecture. The network is constructed to produce the same result in any reference frame and to exactly preserve the total number of neutrinos.}
    \label{fig:neural_network}
\end{figure*}

The objective of the ML model is to take the 24 moments of the neutrino radiation field (four-flux of each species) as input and predict the same 24 moments after the fast flavor instability. Throughout this work, it is assumed that the input moments are approximately homogeneous, and the output moments are averaged over the spatial volume of the simulation (which allows to safely overlook the microscopic chaoticity of the system~\cite{Urquilla:2024bvf}). We label the energy-integrated number four-flux moments as $\F^{\alpha n f}$, where $\alpha\in\{t,x,y,z\}$ is a spacetime index, $n\in\{\nu,\bar{\nu}\}$ indexes whether the moment belongs to neutrinos or antineutrinos, and $f\in\{e,\mu,\tau\}$ indexes the neutrino flavor:\footnote{Although we only consider in this paper the 3-flavor case, we will keep some definitions generic by introducing the number of flavors $N_F$.}
\begin{equation}
\label{eq:def_F_fourflux}
    \F^{\alpha \nu f} \equiv \int{\dd p \frac{p^2}{(2 \pi)^3}} \int{\dd \vec{n} \frac{p^\alpha}{p} f_{\nu_f}(p,\vec{n})} \, ,
\end{equation}
with $f_{\nu_f}$ the classical distribution function of $\nu_f$ neutrinos, and similarly for antineutrinos. Using the standard notations $N$ for the number density and $\vec{F}$ for the spatial number flux density, we have for instance, in the comoving frame, $\F^{t \nu e} = N_{e}$, $\F^{t \bar{\nu} \tau} = \bN_\tau$ or $\F^{z \nu \mu} = F_{\mu}^{z}$. The moments are usually chosen to be expressed in the frame comoving with the fluid, since that is where neutrino-matter interactions are most easily defined, but the number four-flux is a relativistic vector.

One of the main design criteria of the ML model is to ensure that it is invariant under rotations and clear to interpret in relativistic settings. To achieve this, we restrict the inputs to be dot products between the pairs of number four-fluxes and between each four-flux and the four-velocity of the fluid, assuming a metric signature of $(-,+,+,+)$. That is, our array of inputs is a set of dot products for each unique pair of $\{(n,f),(m,g)\}$:
\begin{multline}
    \mathsf{X} = \left\{\frac{\F^{\alpha n f}}{N_\mathrm{tot}} u_\alpha\right\}_{n \in \{\nu,\bar{\nu}\},f \in \{e,\mu,\tau\}} \\
    \cup \left\{\frac{\F^{\alpha n f}}{N_\mathrm{tot}} \frac{\F_{\alpha}^{\phantom{\alpha} m g}}{N_\mathrm{tot}}\right\}_{(n,m) \in \{\nu,\bar{\nu}\}, (f,g) \in \{e,\mu,\tau\}}\,\,,
\end{multline}
where the spacetime index is lowered via the metric and $N_\mathrm{tot}=-\sum_{n f}\F^{\alpha n f}u_\alpha$. For 6 species of neutrinos, these are 27 unique quantities.

We use a neural network (NN) whose structure is shown on Fig.~\ref{fig:neural_network}. Between the input and output layers are three fully-connected hidden layers with leaky rectified linear unit (ReLU) activation functions and a leaky slope parameter of 0.01. We find no improvement but increased training duration for deeper networks, and network width is much more effective at improving prediction quality. Each hidden layer has a width of 128, since with our given dataset size, larger widths did not significantly increase prediction quality and were more prone to over-training.

In order to maintain rotational invariance in the output as well, we express the post-instability moments as a linear combination of the input moments
\begin{equation}
\label{eq:transform_ML}
    \F^{\alpha n f \prime} = \yML^{n f}_{m g} \F^{\alpha m g}\,\,.
\end{equation}
The values of the coefficients $\yML^{n f}_{m g}$ are themselves a function of the set of 36 numbers $\widetilde{\yML}^{n f}_{m g}$ that are the direct outputs of the neural network. This extra layer (“Conserve Layer” on Fig.~\ref{fig:neural_network}) enforces conservation of energy and particle number:
\begin{equation}
\label{eq:y_cons_TrN}
    \yML^{n f}_{m g} \equiv \widetilde{\yML}^{n f}_{m g} + \frac{1}{N_F} \left(\delta^n_m - \sum_h \widetilde{\yML}^{n h}_{m g} \right)\,\,.
\end{equation}
When contracted with $\F^{\alpha m g}$, the first term in parentheses separately sums over all neutrino and antineutrino flavors. The second term does the same for $\widetilde{\yML} \cdot \F$, such that the term in parentheses yields the net deficit number current for neutrinos ($n=\nu$) and antineutrinos ($n = \bar{\nu})$. We evenly distribute the deficit over the $N_F$ flavors, which forces particle number conservation without preferring any single flavor. The overall action of the ML model can be seen as a complicated non-linear function $\F \mapsto \yML^{nf}_{mg}(\F)$.

We do not explicitly enforce conservation of lepton number, so the model serves as an honest assessment of the prospects of future models to deal with a more complete set of phenomena that do not respect this symmetry. However, since this particular model is trained on FFI simulations, it should learn to approximately preserve the lepton number of each flavor. This will be tested in the results section.
ELN conservation can of course be enforced after the model's predictions. However, doing so by modifying the $\yML$ tensor leads to conservation of ELN flux, which is not a property of the FFI under periodic boundary conditions, and leads to significant errors (see Appendix~\ref{app:ELN_conservation}).

Although the ML model is fundamentally built for three flavors, we restrict our inputs to have equal distributions of $\nu_\mu$ and $\nu_\tau$ (and separately equal distributions of $\bar{\nu}_\mu$ and $\bar{\nu}_\tau$), reflecting the approximately equal interaction rates of heavy lepton neutrinos in NSM environments. The ML model does learn to predict approximately equal distributions for these neutrino species, but we explicitly average the heavy lepton distributions output by the ML model. This significantly reduces the effective size of the parameter space, allowing the model to converge with many fewer training data points than would be required in general for a three-flavor model. The inputs and outputs of the model could be simplified in this light, but we choose to leave it as a general three-flavor process so it can be more readily extended to other transformation phenomena in the future.

The ML training software is publicly available at Ref.~\cite{RheaCode}.

\subsection{Local \texttt{Emu} Simulations}
\label{subsec:Emu}

Training and testing the ML model requires accurate dynamical simulations of neutrino evolution to determine the asymptotic state of an unstable distribution. Although different choices of boundary conditions and the inclusion of collisional processes can significantly alter the spatial and temporal structure of the FFI (e.g., \cite{Nagakura:2022qko,Nagakura:2022xwe,Zaizen_CharacterizingQuasisteadyStates_2023,Nagakura:2023wbf,Cornelius:2023eop,zaizen_fastswap_2024}), we operate under the assumption that the asymptotic state is determined locally. We use local, centimeter-scale simulations with periodic boundary conditions to estimate the asymptotic state of the FFI. To that end, we use \texttt{Emu}~\cite{richers_ParticleincellSimulationNeutrino_2021}, a particle-in-cell (PIC) code which simulates a large number of computational particles each associated with a position $\vec{x}^{(i)}$, momentum $\vec{p}^{(i)}$, and a $3 \times 3$ flavor density matrix $\vrho_{ab}^{(i)}$ ($\bvrho_{ab}^{(i)}$ for antineutrinos). We perform simulations of the energy-integrated distributions to explore the self-interaction-dominated limit relevant to the FFI.

The PIC implementation of the Quantum Kinetic Equations (QKEs) describing (anti)neutrino dynamics~\cite{sigl_GeneralKineticDescription_1993,blaschke_NeutrinoQuantumKinetic_2016,richers_NeutrinoQuantumKinetics_2019} reads, in the limit adapted to fast flavor conversions, for each particle $(i)$:
\begin{equation}
\label{eq:QKE}
    \begin{aligned}
        \frac{\dd \vrho_{ab}^{(i)}}{\dd t} &= - \frac{\mathrm{i}}{\hbar} \left[\mathcal{H}^{(i)},\vrho^{(i)}\right]_{ab} \, , \\
        \frac{\dd \bvrho_{ab}^{(i)}}{\dd t} &= - \frac{\mathrm{i}}{\hbar} \left[\overline{\mathcal{H}}^{(i)},\bvrho^{(i)}\right]_{ab} \, , \\        
        \frac{\dd \vec{x}^{(i)}}{\dd t} &= c \, \widehat{\vec{p}}^{(i)} \, , \\
        \frac{\dd \vec{p}^{(i)}}{\dd t} &= \vec{0} \, ,\\
        \frac{\dd n^{(i)}}{dt} &= 0 \, ,\\
        \frac{\dd \bar{n}^{(i)}}{dt} &= 0 \, ,       
    \end{aligned}
\end{equation}
where $\widehat{\vec{p}}^{(i)} = \vec{p}^{(i)}/\lvert \vec{p}^{(i)}\rvert$ is the particle's direction, $n^{(i)}$ and $\bar{n}^{(i)}$ are the number of neutrinos and antineutrinos contained in each computational particle, $\mathcal{H}^{(i)}$ is the self-interaction mean-field Hamiltonian reconstructed at ($\vec{x}^{(i)},\vec{p}^{(i)}$), and $\overline{\mathcal{H}}=-\mathcal{H}^*$. The solver uses second-order shape functions and a global fourth-order time integration scheme. The initial flavor diagonal components of the density matrices are attributed to individual particles such that the initial number four-flux $\F$ is reproduced for each flavor, with an angular distribution determined by the classical maximum entropy (ME) closure \cite{Minerbo_1978}, consistently with the choice of closure used in the NSM simulations we will consider. The angular distribution  associated to the moments $(N_a,\vec{F}_a)$ is, in the direction of the unit vector $\vec{n}$:
\begin{equation}
\label{eq:MEC_distrib}
    f_{\nu_a}(\vec{n}) = \frac{N_a}{4 \pi} \frac{Z_a}{\sinh(Z_a)} e^{Z_a (\widehat{\vec{F}}_a\cdot \vec{n})} \, ,
\end{equation}
where $\widehat{\vec{F}}_a \equiv \vec{F}_a/\lvert \vec{F}_a\rvert$, and $Z_a$ is determined by solving:
\begin{equation}
    \frac{\lvert \vec{F}_a\rvert}{N_a} = \coth(Z_a) - \frac{1}{Z_a} \, .
\end{equation}
Therefore, the initial values of $\vrho_{aa}^{(i)}$ and $n^{(i)}$ are set such that $\vrho_{aa}^{(i)}n^{(i)} = (4\pi V_\text{cell}/n_\text{ppc}) f_{\nu_a}(\widehat{\vec{p}}^{(i)})$ and $\Tr\left[\vrho^{(i)}\right]=1$, where $n_\text{ppc}$ is the number of computational particles per cell and $V_\text{cell}$ is the spatial volume of the cell. This allows the moments $(N_a,\vec{F}_a)$ to be reconstructed with enough angular resolution.

\begin{figure}[!tb]
    \centering
    \includegraphics[width=0.9\linewidth]{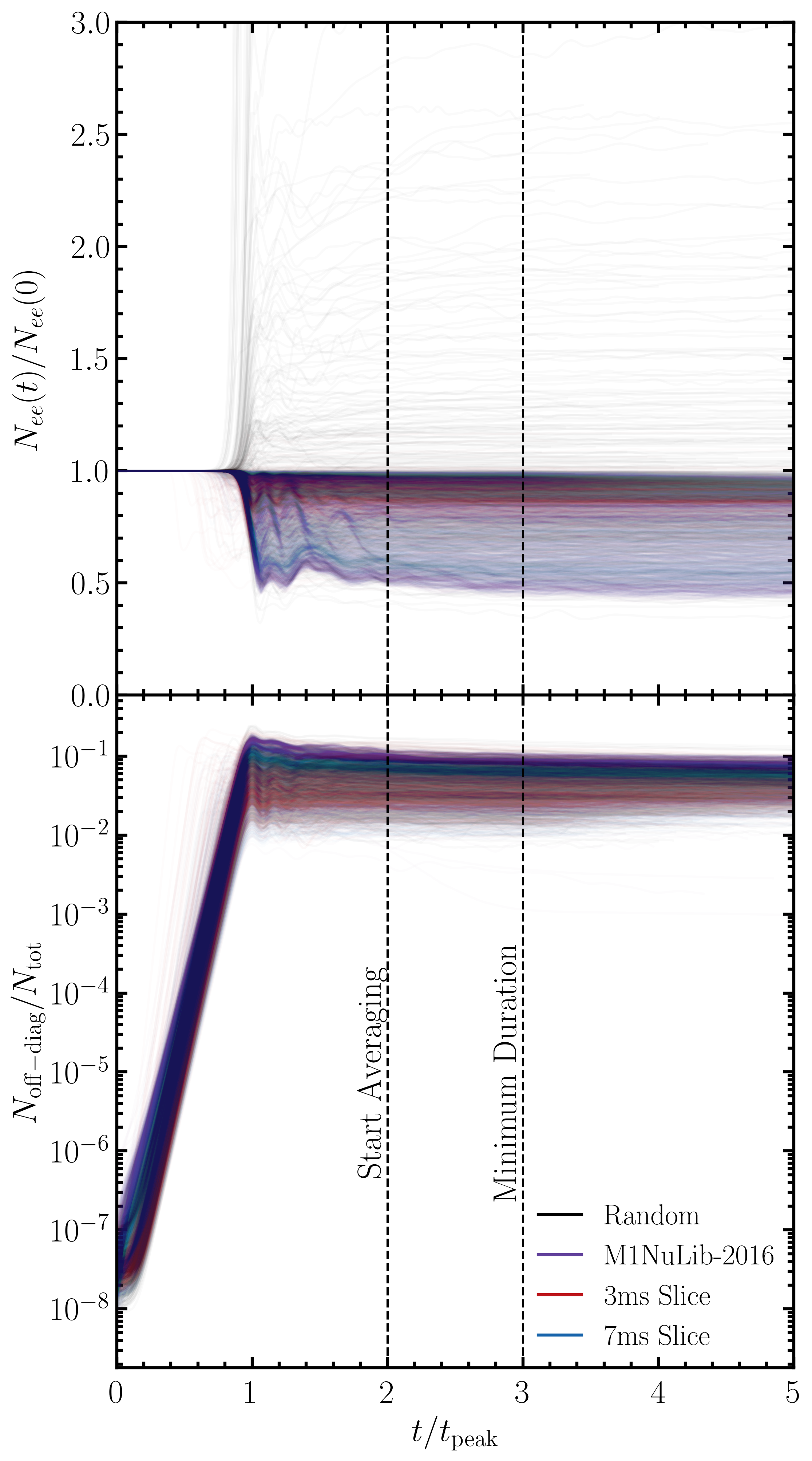}
    \caption{Results of a subset of the \texttt{Emu} simulations representative of the full training dataset (see Sec.~\ref{subsec:training_data}) and data reduction procedure. \emph{Top:} evolution of the electron neutrino number density. \emph{Bottom:} evolution of the flavor off-diagonal number density [see Eq.~\eqref{eq:def_N_offdiag}], showing the exponential growth of the instability until $t=t_\mathrm{peak}$ and the subsequent decoherence phase. The time axis of each simulation is rescaled by $t_\mathrm{peak}$, which corresponds to the saturation of the instability.}
    \label{fig:training_data}
\end{figure}

The fast flavor instability is associated with the development of flavor coherence from an initially almost flavor-diagonal density matrix. In other words, the quantity
\begin{equation}
\label{eq:def_N_offdiag}
    N_\text{off-diag} \equiv \frac{1}{V_\mathrm{domain}}\sum_i n^{(i)}\sqrt{\big\lvert \vrho^{(i)}_{e \mu}\big\rvert^2 + \big\lvert \vrho^{(i)}_{e \tau} \big\rvert^2 + \big\lvert \vrho^{(i)}_{\mu \tau} \big\rvert^2} \, ,
\end{equation}
with $V_\text{domain}$ the domain volume, grows exponentially from $N_\text{off-diag}/N_\text{tot} \ll 1$ to $N_\text{off-diag}/N_\mathrm{tot} = \mathcal{O}(1)$ (the growth rate and associated lengthscale of the instability can be determined by linear stability analysis~\cite{Froustey_LSA_2023}), before entering the non-linear decoherence phase. Examples are shown on Fig.~\ref{fig:training_data}, and the datasets represented by each color are introduced in Sec.~\ref{subsec:training_data}.

All flavor transformation simulations in this work are one-dimensional. Each dataset has different choices regarding the particular simulation resolution, spatial extent, and number of grid cells (see specifics in Sec.~\ref{subsec:training_data}). However, in all cases we first rotate the (anti)neutrino fluxes such that the net ELN-XLN flux lies in the $z$-direction, and also discretize space only in this direction. This artificially enforces homogeneity in the $x$- and $y$-directions, but since the fastest growing unstable mode closely aligns with this direction, we are able to capture the dominant dynamical features. Note that although we assume homogeneity in two directions, we do not assume isotropy or axial symmetry in momentum space.

For the simulations used in the paper, we impose random perturbations to the flavor off-diagonal components of the particle density matrices with magnitudes on the order of $10^{-6}$ relative to the diagonals. We guarantee that each simulation represents a robust FFI by only accepting simulations in which $N_\text{off-diag}$ has a local maxima (at $t=t_\mathrm{peak}$) separated from the initial value by at least 3 orders of magnitude. This  criterion allows us to distinguish between small fluctuating perturbations and genuine FFIs. We then average the final distributions from $t=2 t_\mathrm{peak}$ to the end of the simulation. We ensured that the duration of each simulation is at least 3 times the time required to reach the saturation of the FFI (end of the exponential growth of $N_\text{off-diag}$), as seen in Fig.~\ref{fig:training_data} — this ensures that we are able to average over a time duration of at least $t_\mathrm{peak}$.

\subsection{Training Data}
\label{subsec:training_data}

The ML model is trained on several datasets composed of both stable and unstable input distributions. These include poloidal slices of neutron star merger simulations and randomly generated distributions, which we describe in the following subsections. The data used in training our ML models is publicly available at Ref.~\cite{dataset}.

\subsubsection{NSM Simulation Snapshots}

\begin{figure}[!htb]
    \centering
    \includegraphics[width=\linewidth]{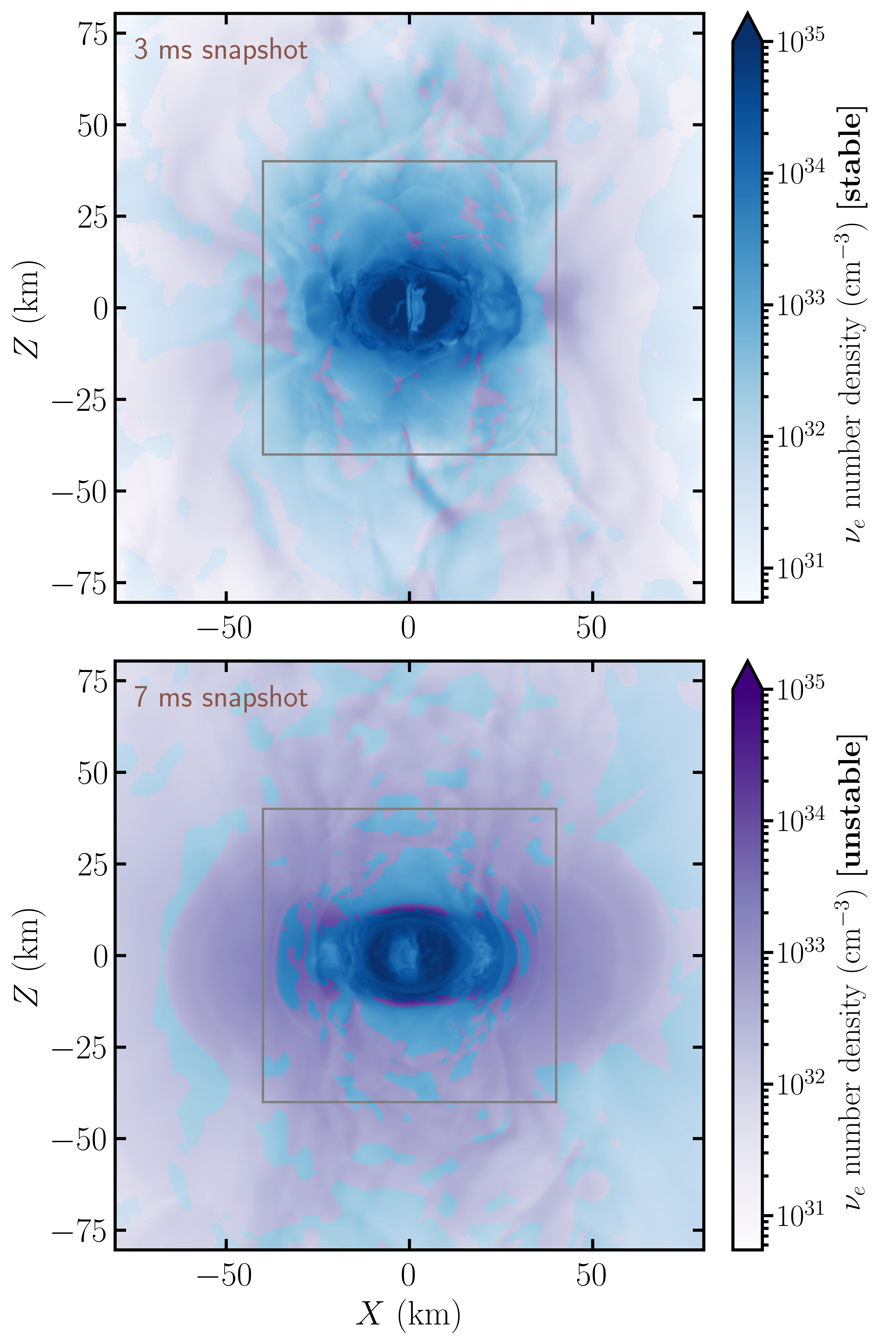}
    \caption{Electron neutrino number density on the transverse slices from each NSM simulation snapshot, used for the training of the ML model. The (purple) blue regions correspond to fast flavor (un)stable points. The unstable points used for the training of the fiducial NN are located outside the inner gray squares in each snapshot.}
    \label{fig:snapshots}
\end{figure}

We use the data from a general relativistic simulation of the merger of two neutron stars with component masses of $1.3 \, M_\odot$ and $1.4 \, M_\odot$~\cite{Foucart:2024npn}. 
The merger simulations are performed with the general relativistic radiation hydrodynamics code SpEC~\cite{Duez:2008rb,Foucart:2012vn}, using the SFHo equation of state for dense matter~\cite{Steiner:2012rk} and a gray two-moment scheme for neutrino transport~\cite{foucart_PostmergerEvolutionNeutron_2015,foucart_ImpactImprovedNeutrino_2016}. Neutrino-matter interaction rates are computed using the NuLib library~\cite{OConnor:2014sgn}, including charged current reactions for electron type (anti)neutrinos, electron-positron annihilations and nucleon-nucleon Bremmstrahlung for heavy-lepton neutrinos, and scattering on neutrons, protons, alpha particles and heavy nuclei for all species (see M1-NuLib simulation in~\cite{Foucart:2024npn}). The remnant is a hypermassive neutron star that collapses to a black hole after $8.5\,\mathrm{ms}$. We consider two pre-collapse snapshots of this simulation here, at $3 \, \mathrm{ms}$ and $7 \, \mathrm{ms}$ post-merger (see Fig.~\ref{fig:snapshots} for two transverse slices).

\begin{figure}[tb]
    \centering
    \includegraphics[width=\linewidth]{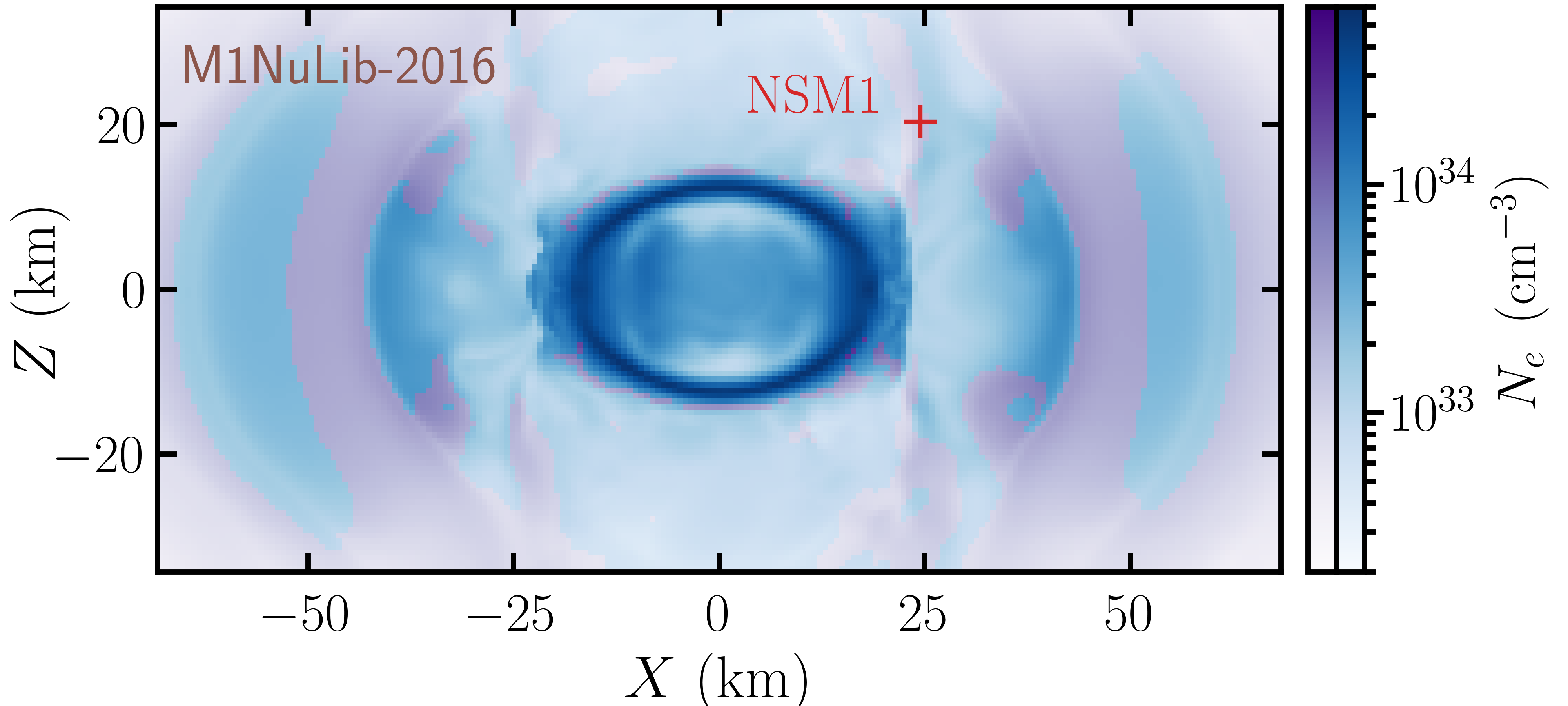}
    \caption{Electron neutrino number density on the transverse slice of the second NSM simulation (“M1NuLib-2016”)~\cite{foucart_ImpactImprovedNeutrino_2016}. As in Fig.~\ref{fig:snapshots}, the (purple) blue points are (un)stable. We identify with a red cross the “NSM1” point~\cite{Grohs_NeutrinoFastFlavor_2022,Grohs:2023pgq} studied in detail in Sec.~\ref{subsec:NSM1}.}
    \label{fig:OldM1_snapshot}
\end{figure}

We also consider the 5-ms post-merger snapshot from a completely independent simulation of a neutron star merger with component masses of $1.2 \, M_\odot$ and $1.2 \, M_\odot$~\cite{foucart_ImpactImprovedNeutrino_2016}, shown on Fig.~\ref{fig:OldM1_snapshot}. This simulation (subsequently referred as “M1NuLib-2016”) is performed with the same two-moment radiation transport code, but with the LS220 equation of state~\cite{Lattimer:1991nc}. In this low mass symmetric system, a larger fraction of the matter is in the neutron star remnant, rather than in the accretion disk or matter outflows. This second system also has a more stable neutron star remnant: the higher mass, asymmetric system previously discussed collapses after $\sim 8\,{\rm ms}$, while this simulation shows no sign of collapse to a black hole for its $10\,{\rm ms}$ of evolution. This snapshot is used as test data to assess the performance of the ML model in a different environment.

The M1 scheme used in these simulations only evolves the first angular moments of the neutrino distributions (number density $N$, flux density $\vec{F}$) for three species: $\nu_e$, $\bar{\nu}_e$ and $\nu_x$ (in a three-flavor framework, this corresponds to assuming $\nu_\mu = \nu_\tau = \bar{\nu}_\mu = \bar{\nu}_\tau \equiv \nu_x$). The system of equations is closed by specifying an analytic form for the pressure tensor as a function of $(N,\vec{F})$, namely, the “Minerbo” closure~\cite{Minerbo_1978,cernohorsky_MaximumEntropyDistribution_1994,smit_ClosureFluxlimitedNeutrino_2000,murchikova_AnalyticClosuresM1_2017}. Without detailed angular distributions, we determine which locations are unstable to the FFI by searching for electron lepton number (ELN) crossings in the maximum entropy distributions following the procedure derived in~\cite{Richers_evaluating_2022}.

The training data from these snapshots consists of both stable and unstable datasets. In the stable dataset, we take all of the FFI-stable distributions on the $X-Z$ slice from the refinement level described in Fig.~\ref{fig:snapshots}, and assume that the final distribution is equal to the initial distribution for each neutrino species. This amounts to 21,360 stable data points in the 3-ms M1-NuLib snapshot, 18,886 stable data points in the 7-ms M1-NuLib snapshot. We also use FFI-stable distributions from all four refinement levels for the same $X-Z$ slice, amounting to 160,647 and 114,895 stable data points from the 3-ms and 7-ms snapshots respectively, together with 10,970 stable data points from the 5-ms M1NuLib-2016 snapshot to train an additional ML model that we call “ML\_ext” in the following. 

In the unstable training dataset, we take all of the FFI-unstable distributions in the refinement level described in Fig.~\ref{fig:snapshots} and simulate their evolution as described in Sec.~\ref{subsec:Emu}. The domain size in the $z$-direction was chosen such that $L_z=10^{1.5}\lambda_\mathrm{max}$, where $\lambda_\mathrm{max}$ is the wavelength of the fastest growing mode as determined by linear stability analysis~\cite{Froustey_LSA_2023}. The grid for each simulation was composed of 1,024 spatial scales, such that the wavelength of the fastest growing mode also spanned approximately $10^{1.5}$ grid cells. The resulting data were only used if the duration of the simulation was at least three times the time required for the FFI to saturate to ensure sufficient time to average over the asymptotic state as described in Fig.~\ref{fig:training_data}. After this selection process, the 3-ms snapshot yielded 9,199 data points and the 7-ms snapshot yielded 7,849 data points. In the ML\_ext model, we also use 7,579 data points similarly simulated based on unstable distributions in the 5-ms M1NuLib-2016 snapshot. It is clear from Fig.~\ref{fig:training_data} that all of the simulations indeed show a clear signature of linear growth and saturation at a stable final state, as indicated by flat curves for $t/t_\text{peak} \gtrsim 2$.

\subsubsection{Randomly Generated Distributions}
\label{subsec:random_training}

We use a number of randomly generated distributions to expand the parameter space explored by the model beyond the conditions present in a NSM simulation evolved assuming classical transport. In all of these distributions, we once again assume that muon and tau flavor neutrinos have the same distributions. However, unlike the NSM-based data, neutrinos and antineutrinos do in general have different distributions — this is notably expected to be the case in general after fast flavor conversion, a situation that will arise if the ML model is embedded in a large-scale simulation.

We first run a set of 4,028 1D \texttt{Emu} simulations starting from randomly-generated initial conditions. In all of these simulations, we choose a domain size of $L_z=1024\,\mathrm{cm}$ and resolve the domain with 8,192 grid cells. This ensures that the smallest unstable wavelengths are resolvable by the grid, and all but extremely large wavelengths fit in the domain. In all of these simulations the net number densities of neutrinos and antineutrinos were set to $2.44\times10^{32}\,\mathrm{cm}^{-3}$, such that the smallest unstable wavelength (in the limit of oppositely-directed neutrino beams) is $2\,\mathrm{cm}$ \cite{chakraborty_SelfinducedNeutrinoFlavor_2016}. The number densities of each species was uniformly sampled between 0 and 1, the densities of tau neutrinos were set to be equal to those of muon neutrinos, and all densities were then scaled by a constant factor to achieve the above net density. The directions of the flux vectors for each species was sampled as $\cos(\theta)=U^2$, where $U$ is a uniformly sampled random number between 0 and 1, such that all distributions have a positive $z$ component of the number flux (since one expects all distributions to be moving in roughly the same direction locally in a NSM). The azimuthal direction of each flux was sampled uniformly, and the flux factor was sampled as $\lvert \vec{F}\rvert / N = 0.9 U$. Only unstable distributions were simulated, and after the selection process described for the NSM simulations and shown in Fig.~\ref{fig:training_data}, this process yielded 2,074 data points.

In addition to randomized unstable distributions, we also train with a large number of randomly-generated distributions that are trivially stable, such that we can assume the asymptotic state is equal to the initial one. At the beginning of training each model, we generate $2\times10^5$ distributions with number densities sampled as above, but all flux factors set to zero (labelled “0 flux factor” in plots). In addition, we sample $2\times10^5$ distributions where only a single species has nonzero density, and has a flux direction uniformly sampled in solid angle and a flux factor uniformly sampled between 0 and 1 (called “1 species” in plots). In this case, if the muon or tau neutrino is selected to have nonzero density, both flavors are set with the same distribution. Finally, we generate an additional $2\times10^5$ distributions with number densities uniformly sampled as above, flux directions uniformly sampled in solid angle, and flux factors sampled as $\lvert \vec{F}\rvert/N=(0.95 U)^{10}$ (called “random” in plots). This exponent is used to cause only about half of the sampled points to be stable (and hence used in training), in order to more fully sample the line between stable and unstable distributions. In this case, we test for stability by constructing a maximum entropy distribution for each species, discretizing into 378 directions, and checking for the presence of an ELN crossing. Finally, we generate $2\times10^5$ distributions \textit{at every training epoch} to be used to probe for unphysical predictions. These are generated using the same method as the “random” stable generated points above, but leaving in both the stable and unstable points (labelled “unphysical” in plots).

\subsection{Training Process}
\label{subsec:training_process}

The ML model is trained to minimize the loss function given by \begin{equation}
\begin{aligned}
    \mathcal{L} &= \mathcal{L}_\text{flux direction} +\mathcal{L}_\text{flux factor}+ \mathcal{L}_\text{number density}   \\&+\mathcal{L}_\text{NSM stable} +\mathcal{L}_\text{1 species stable} + \mathcal{L}_\text{0 flux factor stable}\\ & +\mathcal{L}_\text{random stable} + 100 \mathcal{L}_\text{unphysical} \, .
    \end{aligned}
\end{equation}
The first three loss functions are applied to all of the unstable distributions described in Sec.~\ref{subsec:training_data}. Each of the other terms is applied to one of the unique data sets described in Sec.~\ref{subsec:training_data}. 

The directional loss term is designed to contribute nothing to the loss when the predicted and true fluxes lie in the same direction:
\begin{equation}
    \mathcal{L}_\text{flux direction} = \frac{1}{6 n_\mathrm{data}}\sum_{i=1}^{n_\mathrm{data}} \sum_{s=1}^6 \left(1-\frac{\mathbf{F}^\prime_\mathrm{pred} \cdot \mathbf{F}^{\prime}_\mathrm{true}}
    {N^\prime_\mathrm{pred} \, N^{\prime}_\mathrm{true}}\right) \, ,
\end{equation}
where “pred” refers to the moments predicted by the ML model, and “true” to the value obtained in the \texttt{Emu} simulations. To lighten the notation, the quantities in the sum are implicitly evaluated for the data point $i$ and the species $s$ (for instance, if $s=\bar{\nu}_\mu$, $N'_\mathrm{pred}$ is $\F^{t \bar{\nu} \mu \prime}_\mathrm{pred}$). In this work, we assume three neutrino flavors, or six neutrino species including neutrinos and antineutrinos. 

The loss from the predicted flux factor is
\begin{equation}
    \mathcal{L}_\text{flux factor} = \frac{1}{6 n_\mathrm{data}}\sum_{i=1}^{n_\mathrm{data}} \sum_{s=1}^{6} \left(\frac{|\mathbf{F}^\prime_\mathrm{pred}|}{N^\prime_\mathrm{pred}} - \frac{|\mathbf{F}^{\prime}_\mathrm{true}|}
    {N^{\prime}_\mathrm{true}}\right)^2\, ,
\end{equation}
and the number density loss function reads
\begin{equation}
    \mathcal{L}_\text{number density} = \frac{1}{6n_\mathrm{data}}\sum_{i=1}^{n_\mathrm{data}} \sum_{s=1}^6 \frac{\left(N^\prime_\mathrm{pred}-N^\prime_\mathrm{true}\right)^2}{N_\mathrm{tot}^2} \, .
\end{equation}
The unphysical loss function penalizes any predictions that produce negative number densities or flux factors larger than one. Specifically,
\begin{align}
    &\mathcal{L}_\text{unphysical} = \frac{1}{6n_\mathrm{data}}\sum_{i=1}^{n_\mathrm{data}} \frac{1}{N_\mathrm{tot}^2} \\ &\sum_{s=1}^6 \left[ \mathrm{min}(N'_\mathrm{pred},0)^2  + \mathrm{max}\left(|\vec{F}^\prime_\mathrm{pred}|^2-(N^{\prime}_\mathrm{pred})^2,0\right)\right] \, . \nonumber
\end{align}
Finally, the remaining loss functions are computed as
\begin{equation}
\label{eq:loss_other}
    \mathcal{L} = \frac{1}{24 n_\mathrm{data}} \sum_{i=1}^{n_\mathrm{data}} \sum_{s=1}^{6} \sum_{\alpha=t,x,y,z} \frac{\left(\mathcal{F}^{\alpha \prime}_\mathrm{pred}-\mathcal{F}^{\alpha\prime}_\mathrm{true}\right)^2}{N_\mathrm{tot}^2}\, .
\end{equation}

We randomly select 10\% of each dataset to serve as test data, and the remaining 90\% is the data on which the model is optimized. We train the model to minimize the loss function using the AdamW optimizer~\cite{Loshchilov:2017bsp} with an initial learning rate of $2\times10^{-4}$ and a weight decay of 0.01. We also employ a plateau learning rate scheduler with a patience of 500 iterations, a cooldown of 500 iterations, and a reduction factor of 0.5. The reduction in learning rate mitigates large training performance oscillations that arise otherwise. We also use a dropout probability of 0.1 to prevent over-training and improve model generalizability. Larger dropout probabilities tended to add significant noise to the training process that resulted in larger final loss values, and models trained without dropout were easily over-trained. These hyperparameters produced the highest quality models among many iterations of different hyperparameters, and were trained until performance stopped improving. The models are likely constrained by the amount and quality of the input data, but increasing the amount of data generally requires more training time, and despite our data cleaning procedures, the results of the \texttt{Emu} simulations are accurate only to about 1\%. Training higher-quality models could benefit from a larger amount of more precise data.

\begin{figure*}
    \centering
    \includegraphics[width=\linewidth]{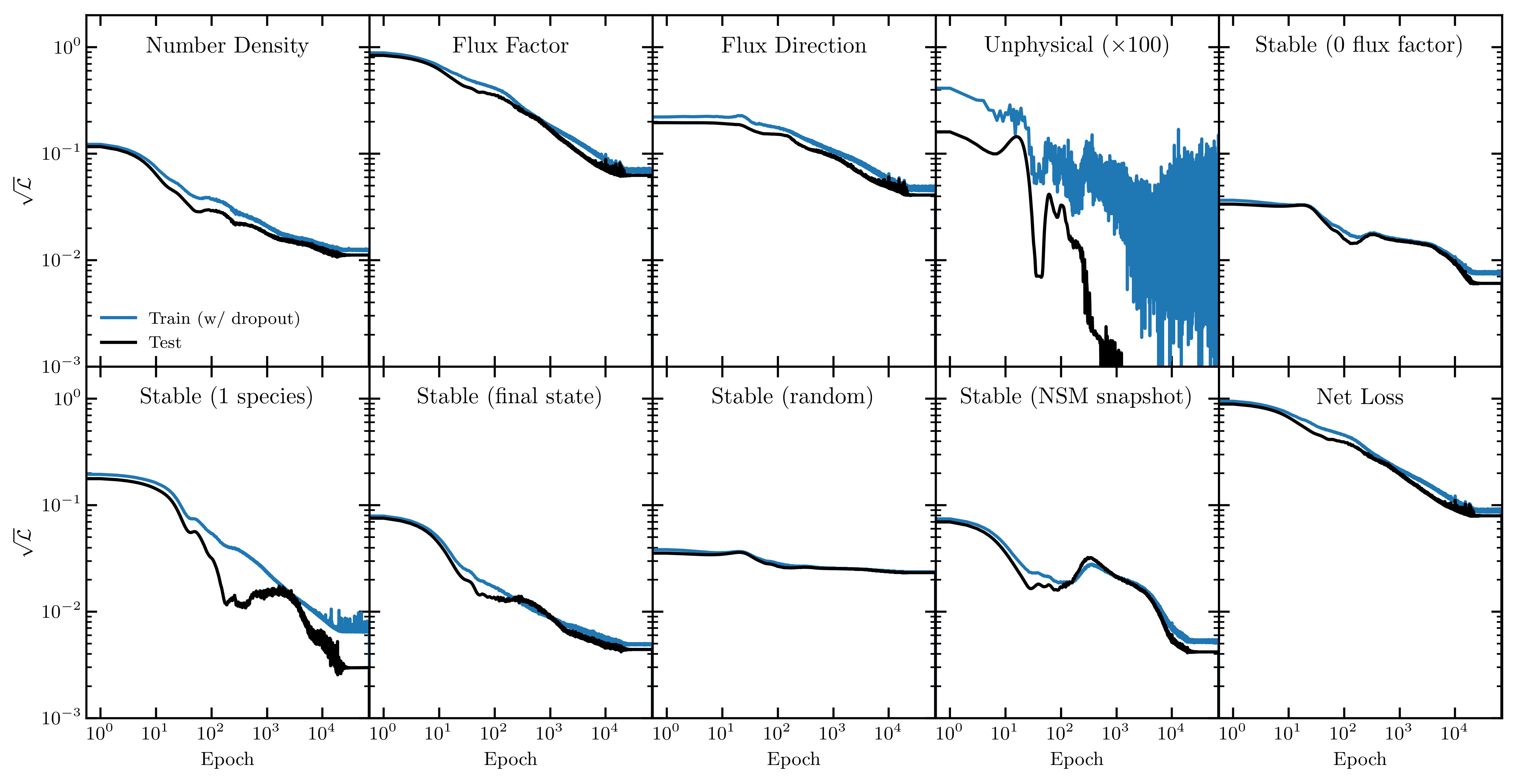}
    \caption{Evolution of the training and test errors over the training of our fiducial ML model. The meaning of each metric is described in Sec.~\ref{subsec:training_process}. The blue curve shows the training loss at each epoch, using the network temporarily truncated by the dropout algorithm. The black points show the loss from the full model on the test dataset. The training improves up to about 20,000 epochs before asymptoting.
    }
    \label{fig:training_error}
\end{figure*}

Figure~\ref{fig:training_error} shows how the training and test errors evolve with training epochs. The first thing to note is that the training error is usually above the test error as a result of using dropout. With a dropout probability of 0.1, during each training epoch 10\% of the nodes are removed from the network to force redundancy in the model, so the predictive capabilities are worse than the full model employed to calculate the test loss. The loss decreases very quickly at first, but tapers (notice the logarithmic horizontal axis). After about 20,000 epochs, the model is unable to improve, driving the learning rate down to essentially zero. The model learns to avoid unphysical results within the first thousand epochs or so (black curve in the “Unphysical” panel), although the reduced network used during each training step is not as good at doing so (blue curve in the same panel). We plot the square root of the loss function so that the plotted values roughly represent the magnitude of the error one can expect in each quantity. By the end of the training process, all of the metrics are accurate to within a few percent.

It is reasonable to expect that the final state predicted by the ML model be stable to further transformation, since the FFI works to remove ELN-XLN crossings~\cite{nagakura_TimeDependentQuasisteadyFeatures_2022,Nagakura:2022xwe,Zaizen:2022cik}. However, this is not in general the case in our model due to a combination of model imperfection and the generally faulty assumption that the input distributions follow maximum entropy profiles [Eq.~\eqref{eq:MEC_distrib}]. Although the final state of the \texttt{Emu} simulations indeed exhibits no ELN-XLN crossing, if this distribution is integrated into two moments and redrawn as a maximum entropy distribution, there is no guarantee that the ELN-XLN is still zero in the “crossed” region. Thus, if the output of the ML model (which reflects the arbitrary angular distribution simulated by \texttt{Emu}) is mapped back into the input (which assumes a maximum-entropy distribution), the model is unable to enforce that this new distribution is stable even if the training were perfect. The stability of the final state is demonstrated in the “Stable (final state)” panel of Fig.~\ref{fig:training_error}, using Eq.~\eqref{eq:loss_other} between the moments resulting from the ML model being applied once and twice. As the model learns, it indeed predicts that successive predictions exhibit smaller changes, but it never goes to zero. We come back to this issue in Sec.~\ref{subsec:NSM1} on a particular example.

Overall, the ML model is able to learn trends reasonably well from a wide range of stable and unstable distributions using only their angular moments. We will spend much of the remaining text quantifying this performance on unseen data with respect to other proposed models.

\section{Approximate mixing schemes for fast-flavor conversion}
\label{sec:mixing_schemes}

Since the computational cost of local QKE calculations is prohibitive in large-scale simulations, finding other ways to determine the asymptotic state reached by the system after the FFI is a necessity. 

ML models are an attractive and efficient way to determine this final state: once trained, the use of an ML model is very fast, and the training can be straightforwardly adapted to other flavor conversion mechanisms. But there have also been many other studies focused on finding analytical prescriptions for the asymptotic state that can also be efficiently implemented in a large-scale simulation. In this section, we introduce a few of the analytical schemes previously designed in the literature, along with a new, fully three-dimensional scheme extended from~\cite{Zaizen:2022cik,Zaizen_CharacterizingQuasisteadyStates_2023}, so we can compare their performance with the ML model in the next section.

\subsection{Existing Mixing Schemes}

The approximate schemes used in the literature are usually built assuming identical distributions of $\mu$ and $\tau$ flavors, called in the following $\nu_x$ for neutrinos and $\bar{\nu}_x$ for antineutrinos. As previously, we will denote with a prime the quantities in the post-FFI asymptotic state.

A range of prescriptions are designed directly at the level of the angular moments, and amount to some sort of flavor equilibration constrained by the symmetries of the problem (e.g., ELN conservation). In the following, we consider several such schemes from Ref.~\cite{just_FastNeutrinoConversion_2022}.

\subsubsection{“Mix1”}

The “Mix1” case assumes explicit conservation of the ELN and XLN (the latter being zero in the NSM points we consider), with equipartition between the species with the lowest number densities. The equipartition value is
    \begin{equation}
        N_\mathrm{eq} = \frac{1}{3} \mathrm{min}\left\{N_e, \bN_e\right\} + \frac{2}{3} \mathrm{min}\left\{N_x, \bN_x\right\} \, ,
    \end{equation}
    and the final number densities are set to $\mathrm{min}\left\{N_e, \bN_e\right\} \to N_\mathrm{eq}$, $\mathrm{min}\left\{N_x, \bN_x\right\} \to N_\mathrm{eq}$, the remaining number densities being determined by ELN and XLN conservation. Note that the total number of neutrinos and antineutrinos is, as expected, conserved with this prescription. These conditions can be, in practice, rewritten in a form similar to Eq.~\eqref{eq:transform_ML}:
\begin{equation}
\label{eq:mix1}
    \F^{tnf \prime} = \left[\yML_\mathrm{Mix1}\right]^{nf}_{mg} \F^{t m g} \, .
\end{equation}
    The fluxes are modified in the same way, that is, $\F^{j n f '} = \left[\yML_\mathrm{Mix1}\right]^{nf}_{mg} \F^{j m g}$ for $j \in \{x,y,z\}$.

\subsubsection{“Mix1f”}

The “Mix1f” case assumes the same transformation for the number densities as Mix1, but conserves flux factors instead of applying the same transformation matrix to the fluxes:
    \begin{equation}
        \F^{j n f \prime} = \left(\frac{\F^{j n f}}{\F^{t n f}}\right) \F^{t n f \prime} \, , \ j \in \{x,y,z\} \, .
    \end{equation}
    A similar scheme was considered in Ref.~\cite{Ehring_FastNeutrinoFlavor_2023}, where the flux factors are conserved for the species whose population actually decrease with flavor conversion, the remaining fluxes being set to conserve the total flux, but in this work we use the version in \cite{just_FastNeutrinoConversion_2022}.

\subsubsection{“Mix2”}

The equipartition treatment of Ref.~\cite{li_NeutrinoFastFlavor_2021} (called “Mix2” in~\cite{just_FastNeutrinoConversion_2022}) does not conserve separately the ELN and XLN, but instead the total lepton number (ELN + XLN). It assumes separate equipartition of neutrinos and antineutrinos:
    \begin{equation}
        \begin{aligned}
            N'_e = N'_x &= \frac13 N_e + \frac23 N_x \, , \\
            \bN'_e = \bN'_x &= \frac13 \bN_e + \frac23 \bN_x \, .
        \end{aligned}
    \end{equation}
    In terms of the $\yML$ transformation matrix, it reads for 3 flavors:
    \begin{equation}
        \left[\yML_\mathrm{Mix2}\right]^{nf}_{mg} = \delta^{n}_{m} \times \frac{1}{3} \begin{pmatrix} 1 & 1 & 1 \\ 1 & 1 & 1 \\ 1 & 1 & 1 \end{pmatrix} \, .
    \end{equation}
    The fluxes are modified with the same mixing coefficients. This scheme breaks the symmetry $N'_x = \bN'_x$, and in general violates net ELN conservation that should be enforced by the QKEs in the FFI limit. This scheme is thus expected to give less accurate results compared to the Mix1(f) cases.

\subsubsection{Survival Probability Prescription}

Other works have not aimed at determining directly the angular moments post-FFI, but rather the full angular distributions of neutrinos. Detailed studies have thus led to the design of a prescription for the direction-dependent survival probability~\cite{bhattacharyya_FastFlavorDepolarization_2021,Bhattacharyya_ElaboratingFate_2022,nagakura_TimeDependentQuasisteadyFeatures_2022,Zaizen:2022cik,Zaizen_CharacterizingQuasisteadyStates_2023,Xiong:2023vcm,Abbar_ML_outcome}, i.e., the probability for an electron neutrino to still be an electron neutrino after the FFI. For axially symmetric distributions, and assuming identical initial distributions of $\nu_\mu,\nu_\tau,\bar{\nu}_\mu,\bar{\nu_\tau}$ (that is, $\text{XLN}=0$), these prescriptions generally enforce full flavor equipartition on the “small side” of the angular ELN crossing, i.e., the angular range where the difference between $\nu_e$ and $\bar{\nu}_e$ distributions is smaller in absolute value. The survival probability on the “large side” is then set by ELN conservation with a prescribed angular dependence. This probability is for instance described as a box-like function (which creates artificial discontinuities in the final angular distributions)~\cite{Zaizen:2022cik,Zaizen_CharacterizingQuasisteadyStates_2023} or with a continuous transition~\cite{Xiong:2023vcm,Abbar_ML_outcome,Xiong:2024pue}. Since it was shown to be superior to the box-like form in axisymmetric simulations, we consider the latter scheme, with the so-called “Power-1/2” analytic form (see details in~\cite{Xiong:2023vcm}).

These prescriptions require knowledge of the angular distributions of (anti)neutrinos. Since we are applying these schemes to distributions described only by moments, we assume angular distributions given by the classical maximum entropy closure~\eqref{eq:MEC_distrib}, assuming axisymmetry around the direction of the net ELN-XLN flux. In practice, we set $F^x=F^y=0$ after rotating the system to place the net ELN-XLN flux in the $\hat{z}$ direction.
The final distributions calculated accordingly to the Power-1/2 mixing scheme are then integrated following Eq.~\eqref{eq:def_F_fourflux} to obtain the predicted moments.

\subsection{Three-Dimensional, Non-Axisymmetric Mixing Scheme}

The box-like scheme of~\cite{Zaizen:2022cik,Zaizen_CharacterizingQuasisteadyStates_2023} can be straightforwardly generalized for non-axisymmetric distributions in the following way.\footnote{We assume once again equal distributions for heavy-lepton flavor neutrinos ($f_{\nu_\mu}=f_{\nu_\tau} \equiv f_{\nu_x}$), and heavy-lepton flavor antineutrinos ($f_{\bar{\nu}_\mu}=f_{\bar{\nu}_\tau}\equiv f_{\bar{\nu}_x}$).} We define the ELN-XLN distribution in a given direction $\vec{n}$ by $G(\vec{n}) \equiv \left[f_{\nu_e}(\vec{n}) - f_{\bar{\nu}_e}(\vec{n})\right] - \left[f_{\nu_x}(\vec{n}) - f_{\bar{\nu}_x}(\vec{n})\right]$.
The fast flavor instability is associated to the existence of a crossing in $G$~\cite{Morinaga_Crossing_2022,dasgupta_CollectiveNeutrinoFlavor_2022}, which means that the angular space can be divided in two regions:
\begin{subequations}
\begin{align}
    I_+ &= \left\lvert\int{\mathrm{d}{\vec{n}} \, G(\vec{n}) \times \Theta[G(\vec{n})]}\right\rvert \, , \\
    I_- &= \left\lvert\int{\mathrm{d}{\vec{n}} \, G(\vec{n}) \times \Theta[-G(\vec{n})]}\right\rvert \, ,
\end{align}
\end{subequations}
where $\Theta$ is the Heaviside distribution. Let us denote by $I_<$ ($I_>$) the smallest (largest) value between $I_+$ and $I_-$, and $\Omega_<$ ($\Omega_>$) the associated domain of integration. We then define the survival probability:
\begin{equation}
    \mathbb{P}(\vec{n}) = \left\{ \begin{aligned}
        &\frac{1}{3} &&\text{for } \vec{n} \in \Omega_< \, , \\
        &1 - \frac{2 I_<}{3 I_>} &&\text{for } \vec{n} \in \Omega_> \, .
    \end{aligned} \right.
\end{equation}
The final angular distributions are set to:\footnote{Since we are considering pairwise conversions between $(\nu_e,\bar{\nu}_e)$ and $(\nu_{\mu,\tau},\bar{\nu}_{\mu,\tau})$, we have transitions $\nu_a \to \nu_a$ with probability $\mathbb{P}$, transitions $\nu_a \to \nu_b$ ($a \neq b$) with probability $(1-\mathbb{P})/2$, and similarly for antineutrinos. Accounting for $\nu_\mu = \nu_\tau = \nu_x$ and $\bar{\nu}_\mu = \bar{\nu}_\tau = \bar{\nu}_x$ leads to \eqref{eq:final_Box3D}.}
\begin{equation}
\label{eq:final_Box3D}
    \begin{aligned}
        f_{\nu_e}'(\vec{n}) &= \mathbb{P}(\vec{n}) f_{\nu_e}(\vec{n}) + \left[1 - \mathbb{P}(\vec{n})\right] f_{\nu_x}(\vec{n}) \, , \\
        f_{\bar{\nu}_e}'(\vec{n}) &= \mathbb{P}(\vec{n}) f_{\bar{\nu}_e}(\vec{n}) + \left[1 - \mathbb{P}(\vec{n})\right] f_{\bar{\nu}_x}(\vec{n}) \, , \\
        f_{\nu_x}'(\vec{n}) &= \frac12 \left[1-\mathbb{P}(\vec{n})\right] f_{\nu_e}(\vec{n}) + \frac12\left[1 + \mathbb{P}(\vec{n})\right] f_{\nu_x}(\vec{n}) \, , \\
        f_{\bar{\nu}_x}'(\vec{n}) &= \frac12 \left[1-\mathbb{P}(\vec{n})\right] f_{\bar{\nu}_e}(\vec{n}) + \frac12\left[1 + \mathbb{P}(\vec{n})\right] f_{\bar{\nu}_x}(\vec{n}) \, ,
    \end{aligned}
\end{equation}
with $f_{\nu_\mu}' = f_{\nu_\tau}' = f_{\nu_x}'$ and likewise for antineutrinos. We dub this procedure “Box3D.” As in the Power-1/2 scheme, the final state is discretely integrated to obtain the final four-flux $\F'$. Note that reproducing the initial moments with higher precision require more angular bins, hence a higher computation time (compared to the ML model, for instance). Although the Power-1/2 scheme is justified by simulations of axisymmetric systems, there has not been to date systematic studies of non-axisymmetric configurations that would support the Box3D generalization we propose here. The results of Sec.~\ref{sec:performance} will nevertheless show that this model offers the best overall performance in a range of environments, making it a strong candidate in non-axisymmetric cases.

\paragraph*{Symmetry conservation —} The solution of the QKEs~\eqref{eq:QKE} in the fast flavor limit must preserve a number of symmetries. With periodic boundary conditions, the following quantities (averaged over the simulation box) must be conserved: $\Tr(N)$, $\Tr(\bN)$, $\mathrm{ELN}=N_e-\bN_e$, $\mathrm{XLN}=N_\mu-\bN_\mu = N_\tau - \bN_\tau$, $\Tr(\vec{F})$, $\Tr(\vec{\bF})$. We summarize in Table~\ref{tab:mix_methods} how the different subgrid models satisfy these conservation requirements.

\renewcommand{\arraystretch}{1.3}

\begin{table}[!ht]
    \begin{tabular}{|r|c|c|c|c|c|c|}
    \hline
          & $\mathrm{Tr}(N)$ & $\mathrm{Tr}(\bN)$ & ELN & XLN & $\mathrm{Tr}(\vec{F})$ & $\mathrm{Tr}(\vec{\bF})$  \\
         \hline
         Mix1 & Yes & Yes & Yes & Yes & Yes & Yes \\
         Mix1f & Yes & Yes & Yes & Yes & No & No  \\
         Mix2 & Yes & Yes & No & No & Yes & Yes  \\
         Power-1/2 & Yes & Yes & Yes & Yes & \textit{Yes} & \textit{Yes}  \\
        Box3D & Yes & Yes & Yes & Yes & Yes & Yes \\ \hline
    \end{tabular}
    \caption{\label{tab:mix_methods} Conservation of a few quantities, which are conserved by the exact QKEs with periodic boundary conditions, in the various mixing schemes existing in the literature. Note that the “Power-1/2” scheme is only valid for axisymmetric distributions (hence italic font), so that fluxes are conserved only if they are all parallel to $\hat{z}$.}
\end{table}

The Box3D scheme conserves by construction all these quantities, but its actual implementation relies on an angular discretization of momentum space. As a consequence, for any given angular resolution, the initial four-flux $\F$ is reconstructed with a given precision. The conserved quantities $\Tr(N)$, etc. are then conserved, but with the systematic error associated to the initial discretization. In this paper, we use 100 angular bins for $\theta \in [0,\pi]$ and likewise for $\varphi \in [0, 2 \pi]$, which leads to negligible resolution errors.

\section{Subgrid Model Performance}
\label{sec:performance}

Given the \texttt{Emu} simulations introduced in Sec.~\ref{subsec:training_data}, we have at our disposal several datasets on which we can assess the performance of the previous subgrid models. We recall that the ML model is trained on unstable points from the 3-ms and 7-ms snapshots only ; the snapshot of a previous merger model (“M1NuLib-2016”) thus allows to check its capabilities in a different NSM environment.


\begin{figure*}
    \centering
    \includegraphics[width=\linewidth]{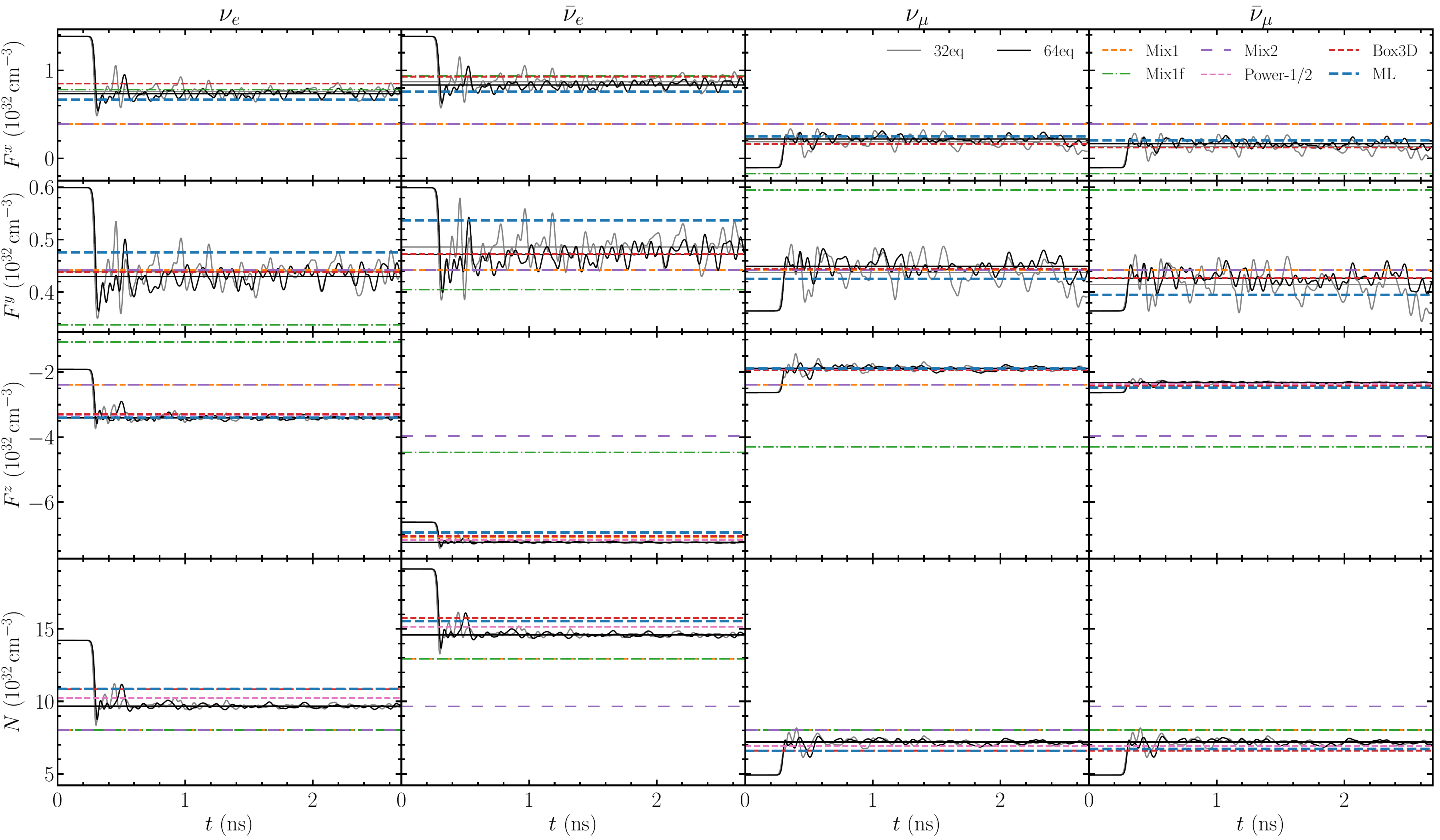}
    \caption{Evolution of the number and flux densities from the conditions of the NSM1 point studied in~\cite{Grohs_NeutrinoFastFlavor_2022,Grohs:2023pgq} (see parameters in Table~\ref{tab:NSM1}), for \texttt{Emu} simulations with 32 equatorial directions (gray, 378 particles per cell) and 64 equatorial directions (black, 1506 particles per cell). Averaging over the final fluctuations (see Fig.~\ref{fig:training_data} and the associated text) leads to the asymptotic values shown by solid horizontal gray/black lines. They can be compared with the predictions of the various mixing schemes introduced in Sec.~\ref{sec:mixing_schemes} and of the ML model, reported as dashed horizontal lines. The moment for $\nu_\tau$ and $\bar{\nu}_\tau$ are not plotted, as they are respectively identical to $\nu_\mu$ and $\bar{\nu}_\mu$.}
    \label{fig:NSM1}
\end{figure*}

\subsection{Performance Metrics}
\label{subsec:performance_metrics}

We assess the accuracy of a prediction of the final four-flux $\F'_\mathrm{pred}$, compared to the true value $\F'_\mathrm{true}$ obtained from a resolved \texttt{Emu} simulation, via the following metrics. Each metric yields a number for a single data point (e.g., one number per location in the NSM snapshot).

First, we assess the accuracy of number density predictions through the maximal normalized difference of number densities,
    \begin{equation}
    \label{eq:err_N}
        \delta_N \equiv \max_{\{n,f\}}{\frac{\left \lvert (\F'_\mathrm{pred})^{tnf} -(\F'_\mathrm{true})^{tnf}\right\rvert}{N_\mathrm{tot}}} \, .
    \end{equation}
Then, the direction of (anti)neutrino fluxes is measured via the maximal misalignment of the spatial flux densities, weighted by the flux factors,
    \begin{equation}
    \label{eq:err_F}
        \delta_{\vec{F}} \equiv \max_{\{n,f\}}\left[\frac{1}{2}\left(1- \frac{\vec{F}'_\mathrm{pred}\cdot \vec{F}'_\mathrm{true}}{\lvert \vec{F}'_\mathrm{pred} \rvert \times \lvert \vec{F}'_\mathrm{true} \rvert} \right) \times \frac{\lvert \vec{F}'_\mathrm{true}\rvert}{N'_\mathrm{true}} \right] \, ,
    \end{equation}
    where inside the brackets, we used the shorthand notations $N' = \F^{tnf\prime}$ and $F^{j\prime} = \F^{jnf\prime}$. Note that $\delta_{\vec{F}}$ is weighted by the flux factor of each species. This is to account for the fact that the smaller the flux factor, the less relevant its precise direction.
    
Finally, we measure the absolute difference of flux factors,
    \begin{equation}
    \label{eq:err_f}
        \delta_f \equiv \max_{\{n,f\}} \left\lvert \frac{\lvert \vec{F}'_\mathrm{pred}\rvert}{N'_\mathrm{pred}} - \frac{\lvert \vec{F}'_\mathrm{true}\rvert}{N'_\mathrm{true}} \right \rvert \, ,
    \end{equation}
    with the same shorthand notations as before.

With these definitions, each of these metrics can have values between 0 and 1.

\subsection{Single Prediction Example}
\label{subsec:NSM1}

We first focus on an example point (“NSM1”), identified on Fig.~\ref{fig:OldM1_snapshot} and previously studied by particle-in-cell and moment methods in~\cite{Grohs_NeutrinoFastFlavor_2022,Grohs:2023pgq}. The classical moments for this NSM1 point from~\cite{foucart_ImpactImprovedNeutrino_2016} are given in Table~\ref{tab:NSM1}.

\renewcommand{\arraystretch}{1.2}

\begin{table}[!ht]
    \begin{tabular}{c|c}
    Moment & Value \\
    \hline 
        $N_{e}$ & $1.42195 \times 10^{33} \ \mathrm{cm}^{-3}$ \\
        $\bN_{e}$ & $1.91462 \times 10^{33} \ \mathrm{cm}^{-3}$ \\
        $N_{\mu,\tau} = \bN_{\mu,\tau}$ & $4.91135 \times 10^{32} \ \mathrm{cm}^{-3}$ \\ \hline
        $\vec{F}_{e}/N_{e}$ & $(\phantom{-}0.0975,0.0422, -0.1343)$ \\
        $\vec{\bF}_{e}/\bN_{e}$ & $(\phantom{-} 0.0724,0.0313,-0.3447)$ \\
        $\vec{F}_{\mu,\tau}/N_{\mu,\tau} = \vec{\bF}_{\mu,\tau}/\bN_{\mu,\tau}$ & $(-0.0217,0.0743,-0.5355)$ \\ 
    \end{tabular}
    \caption{\label{tab:NSM1} Classical moments corresponding to the “NSM1” point.}
\end{table}

In Fig.~\ref{fig:NSM1} we show the evolution of the domain-averaged angular moments calculated by \texttt{Emu}, together with asymptotic state predictions from the subgrid models for the NSM1 point. The four columns show results for four different (anti)neutrino species, and each of the four rows corresponds to a spacetime component of the number flux. The gray curves show the evolution of each moment from an \texttt{Emu} simulation performed with the resolution used in our training dataset (378 particles per cell), and the black curve shows the same using a much higher angular resolution (1506 particles per cell). The proximity of the curves demonstrate that resolution in the \texttt{Emu} simulation is not a significant factor. The solid black line shows the averaged asymptotic state. Recall that this simulation is from a dataset that is not used during training, and the point is selected as a well-studied example, and not for one with particularly good or poor performance. Note that in Sec.~\ref{sec:ensemble_performance} we demonstrate that there is a distribution of errors, such that all sub-grid models exhibit in certain cases larger errors than those shown in Fig.~\ref{fig:NSM1}.

The predicted final state from the ML model is shown in dashed blue. While it is similar to the simulated final state, it is clear that the amount of flavor transformation apparent in the number density (bottom row) is under-predicted, at comparable level with the Box3D scheme. The Power-1/2 scheme performs better, but the Mix1 and Mix1f schemes over-predict the amount of transformation. The Mix2 scheme performs very poorly, notably for antineutrinos (a consequence of the imposed equilibration of $\bN_e$ and $\bN_x$, which is not at all the case in the actual evolution).

The ML and Box3D perform reliably well in predicting the asymptotic values of the flux components. The Mix1 scheme produces reasonable results, but the Mix1f scheme induces significant errors, at times with the opposite trend compared to the true solution. For the same reason as for the number densities, the Mix2 scheme induces large errors on antineutrinos. The Power-1/2 scheme does a very good job of predicting the final flux in the $z$-direction, but the scheme is not able to predict the change in the $x$ and $y$ flux components since it only has an axisymmetric interpretation. Note also that the changes in the $x$ and $y$ fluxes are small compared to the $z$ flux and especially the number density, so errors appear enhanced.

Overall, the ML and Box3D methods show the most consistently good performance across all moments. The performances of all subgrid models applied to the NSM1 data point, as measured by the metrics of Eqs.~\eqref{eq:err_N}--\eqref{eq:err_f}, are reported in Table~\ref{tab:NSM1_results}.
 
\begin{table}[!ht]
    \begin{tabular}{|r|c|c|c|}
    \hline
    Mixing model & $\quad \ \delta_N \quad \ $ & $ \ 10 \times  \delta_{\vec{F}}  \ $ & $\quad \  \delta_f \quad \ $ \\
    \hline \hline
        No conversion & $0.0858$ & $0.1619$ & $0.2707$ \\ \hline
        Mix1 & $0.0312$ & $0.0064$ & $0.0555$ \\
        Mix1f & $0.0312$ & $0.1619$ & $0.2707$ \\
        Mix2 &  $0.0931$ & $0.0053$ & $0.0859$ \\
        Power-1/2 & $0.0104$ & $0.0541$ & $0.0382$ \\
        Box3D & $0.0221$ & $0.0015$ & $0.0486$ \\
        ML & $0.0226$ & $0.0005$ & $0.0496$ \\ \hline
    \end{tabular}
    \caption{\label{tab:NSM1_results} Error metrics, introduced in Sec.~\ref{subsec:performance_metrics}, for the NSM1 point calculations.}
\end{table}

\begin{figure}[!htb]
    \centering
    \includegraphics[width=0.83\linewidth]{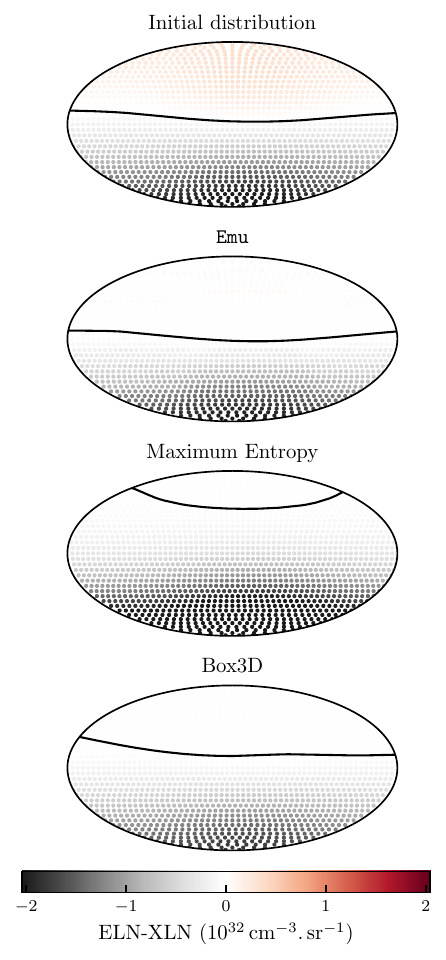}
    \caption{ELN-XLN angular distributions for the NSM1 configuration. From top to bottom: initial state, given by maximum entropy distributions~\eqref{eq:MEC_distrib} for the initial moments (given in Table~\ref{tab:NSM1}); post-FFI asymptotic state distribution from \texttt{Emu}; ELN-XLN based on the maximum entropy distributions built from the final \texttt{Emu} results; final distribution according to the Box3D scheme.}
    \label{fig:final_distribs_NSM1}
\end{figure}

\paragraph*{Asymptotic angular distributions —} As emphasized before, the final angular distributions obtained in \texttt{Emu} are not maximum entropy distributions~\cite{Grohs_NeutrinoFastFlavor_2022,Grohs:2023pgq}. However, embedding our results (through the ML or Box3D model, typically) in a global M1 simulation would, for a given set of final angular moments, a priori implicitly assume that they correspond to a maximum entropy distribution (or another type if a different closure is chosen). We show on Fig.~\ref{fig:final_distribs_NSM1} how the ELN-XLN distribution is affected by this maximum entropy reconstruction.

In the initial state (top panel), we clearly identify the regions of positive (red) and negative (gray) ELN-XLN. The black curve shows the contour of zero ELN-XLN. The final \texttt{Emu} results are in very good agreement with the Box3D scheme: the net ELN-XLN is negative, such that the FFI washes away the ELN-XLN in the angular region where it was initially positive. 
However, looking at the third row (“Maximum Entropy”), where the final moments from the \texttt{Emu} simulation are used to form the equivalent maximum entropy distributions~\eqref{eq:MEC_distrib}, 
we see that the ELN-XLN distribution is significantly different (and a detailed look shows that a positive ELN-XLN part remains, such that an artificial crossing, albeit much smaller, is still present),
which highlights the strong non maximum-entropy-like features of the asymptotic state distributions. Yet, by construction of the M1 scheme with closure, “Maximum Entropy” is the underlying distribution that would be embedded in a global M1 simulation employing any such subgrid model. Solving this inconsistency is beyond the scope of this paper, and is a significant task ahead for quantum moment methods (see also~\cite{Grohs_NeutrinoFastFlavor_2022,Grohs:2023pgq,Froustey_LSA_2023,Froustey:2024sgz}).

\begin{figure*}
    \centering
    \includegraphics[width=0.982\linewidth]{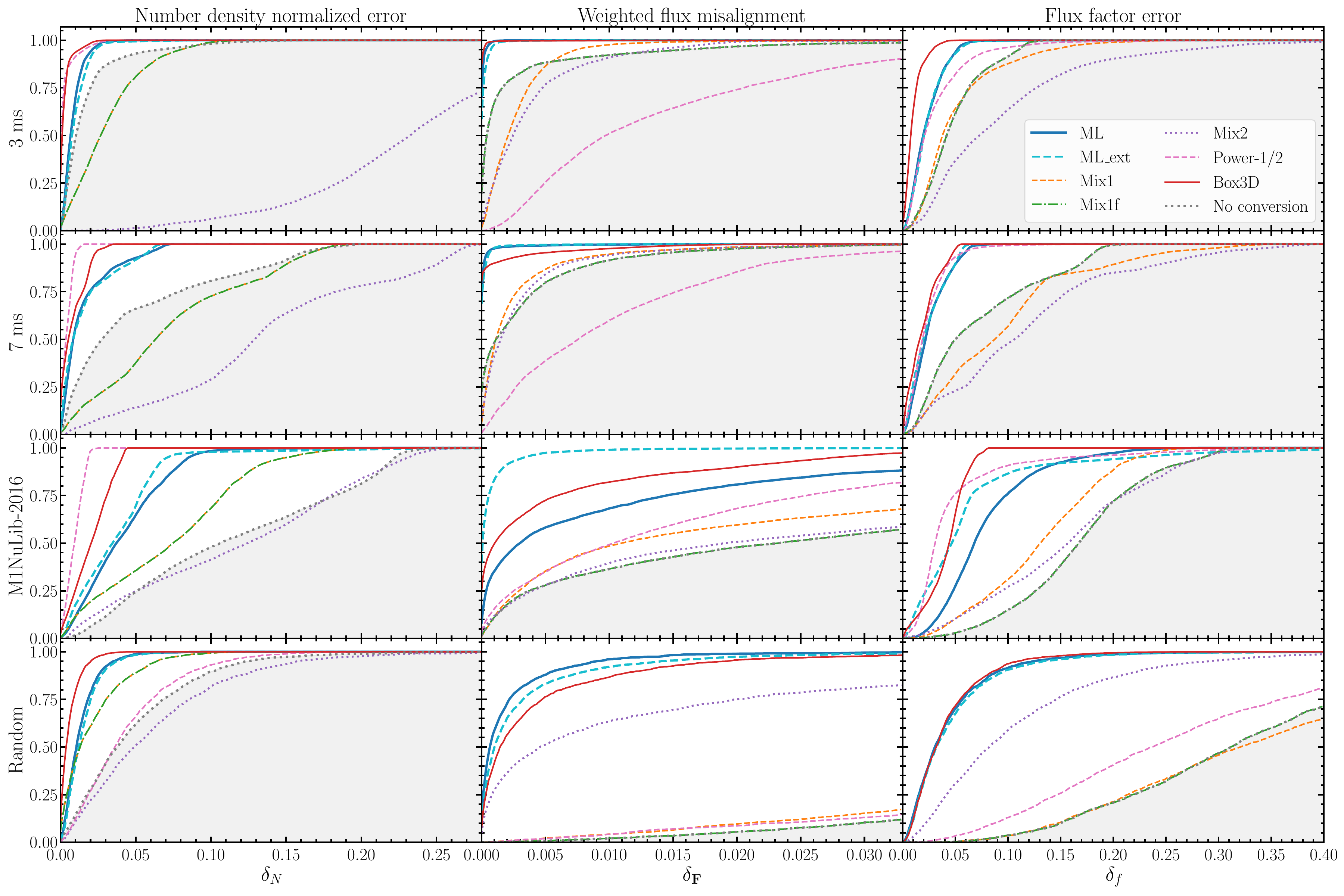}
    \caption{Cumulative distribution functions of the three error metrics [Eqs.~\eqref{eq:err_N}, \eqref{eq:err_F} and \eqref{eq:err_f}] on different datasets of \emph{unstable} points, for the ML model and the different mixing schemes introduced in Sec.~\ref{sec:mixing_schemes}. The dotted gray “No conversion” line shows the error made if no flavor transformation is applied. From top to bottom: 3-ms snapshot from \cite{Foucart:2024npn} (see Fig.~\ref{fig:snapshots}, top panel), 7-ms snapshot from \cite{Foucart:2024npn} (see Fig.~\ref{fig:snapshots}, bottom panel), 5-ms snapshot from \cite{foucart_ImpactImprovedNeutrino_2016} (see Fig.~\ref{fig:OldM1_snapshot}) and random initial states (see Sec.~\ref{subsec:random_training}). Overall good performances are achieved by the Box3D and ML models.}
    \label{fig:CDF_performance}
\end{figure*}

\subsection{Ensemble Performance}
\label{sec:ensemble_performance}
In order to characterize the performance of the models, we apply them to the datasets listed in Sec.~\ref{subsec:training_data} to demonstrate the average behavior and the presence of any outliers. We do not plot results for training and test data separately, since the distributions look very similar. This is in line with the training and test losses plotted in Fig.~\ref{fig:training_error} and indicates that the ML model is not over-trained. Although the model generalizes well to unseen data randomly selected from the same dataset as the training data, we test its ability to generalize to truly new data by testing the model predictions on simulations from an entirely separate simulation of a NSM (“M1NuLib-2016”). We also show the changes in performance if this new simulation is included in the training dataset (“ML\_ext” model, see Sec.~\ref{subsec:training_data}).

\paragraph*{Performance on unstable configurations —} We first demonstrate the ability of each model to correctly predict the outcome of the FFI in \textit{unstable} distributions. In Fig.~\ref{fig:CDF_performance} we show cumulative distribution functions (CDFs) for the amount of error for each performance metric and each set of \texttt{Emu} simulations. The gray dotted curves and the shaded gray area show how incorrect it is to assume that neutrinos do not change flavor. In the 3 ms snapshot, the number densities in unstable locations only change by a few percent (top left panel), indicated by the gray dotted curve quickly rising close to a value of one. The Mix1 and Mix1f schemes over-predict the amount of flavor transformation, but the ML, Power-1/2, and Mix3D schemes perform better. The Mix2 scheme does the poorest job, with large number density errors, which is expected given the incorrect assumptions of equipartition embedded in this model (already discussed in the example of Sec.~\ref{subsec:NSM1}). Looking at the number density errors in the other datasets (i.e., other panels in the left column), this trend is persistent, noting, moreover, that the amount of flavor transformation generally increases when comparing datasets from top to bottom. Power-1/2 does the best job at predicting number densities in the NSM datasets. In the random dataset, which contains highly non-axisymmetric distributions, the assumption of axisymmetry made in deriving the approach is more strongly violated and it performs much more poorly (bottom left panel). In the NSM datasets, the difference in performance on $\delta_N$ between the Power-1/2 and Box3D models is largely due to the discontinuous nature of the Box3D model: the axisymmetric version of the box-like scheme (not shown) is very close to the Box3D curve for $\delta_N$, indicating that an advantage of Power-1/2 comes from the smooth transition between angular regions in its design. Developing a continuous version of the Box3D model is an open question that we do not seek to address in this paper.

In fact, all of the methods that assume axisymmetry do comparatively poorly in both the flux direction and magnitude metrics for the random dataset (bottom center and bottom right panels). The poor performance of the “No conversion” approach demonstrates that in these cases the flux vectors undergo significant changes. Both the ML and Box3D models do surprisingly good at predicting these changes, but it should be pointed out that many of these samples were used to train the ML model, whereas the Box3D model is not informed by any of these simulations.

In the NSM snapshots, the Power-1/2 scheme has reliably small errors for the predicted flux factors (right column), but reliably large errors for the predicted change in flux direction (center column). Once again, the Box3D and ML models provide the most robust performance, but when looking at data the ML model was not trained with (third row), the Box3D model is clearly superior. The dashed light blue line on Fig.~\ref{fig:CDF_performance} shows the performance of the ML\_ext model, which is also trained on the M1NuLib-2016 data. As expected, the results are better compared to the fiducial ML model, in particular for the flux direction (center column). The performance remains inferior to the Box3D model, which might indicate the need for a more intricate ML model than the most simple relativistically covariant form of Eq.~\eqref{eq:transform_ML} to accommodate various NSM environments. All in all, the Box3D mixing scheme seems to describe the outcome of FFI in a large range of situations with a percent-level accuracy.

\begin{figure}[!ht]
    \centering
    \includegraphics[width=\linewidth]{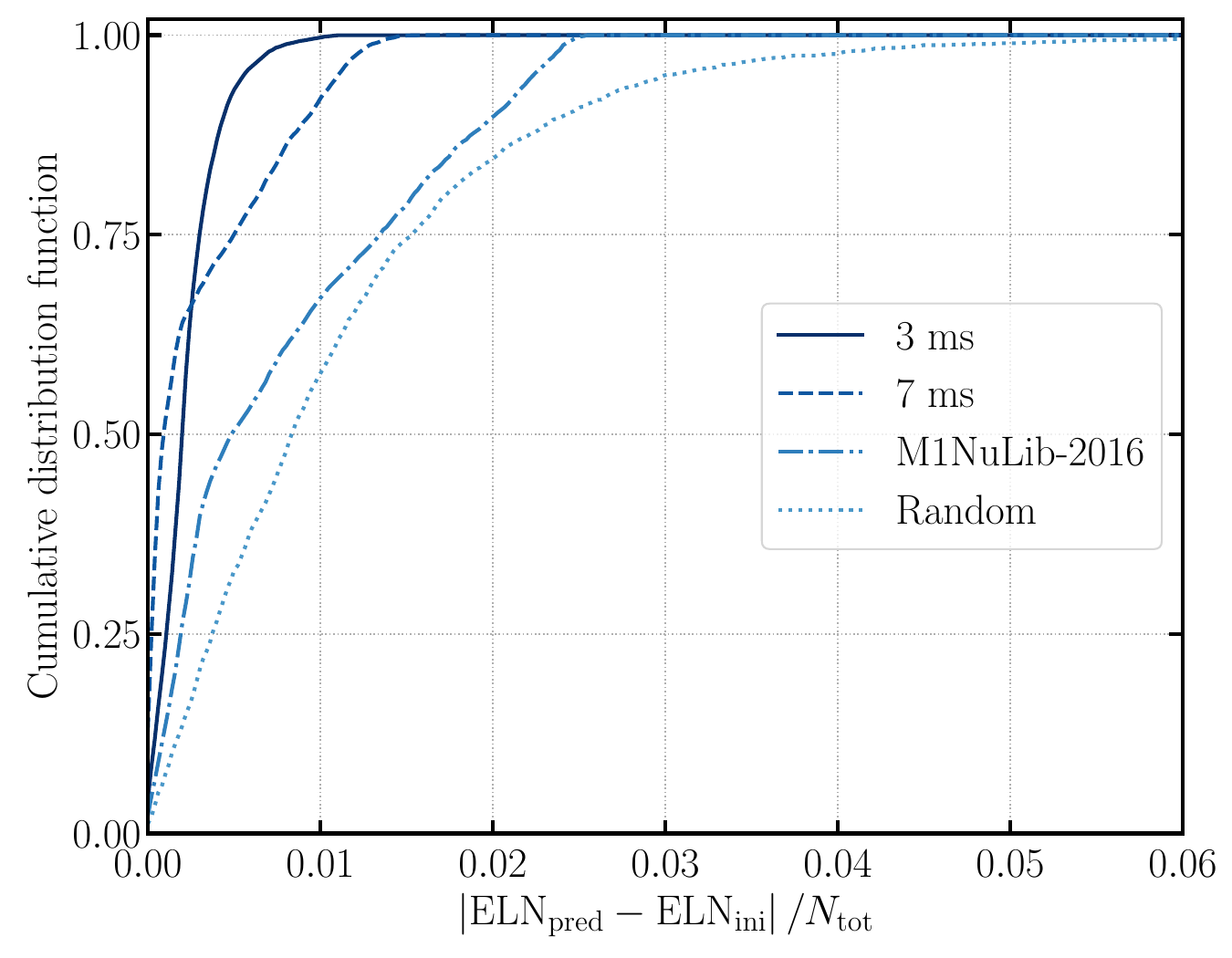}
    \caption{Violation of ELN conservation by the ML model in the various datasets of unstable points.}
    \label{fig:NSM_ELN_violation}
\end{figure}

\paragraph*{Unphysical predictions —} The most egregious errors committed by subgrid models are predictions that violate basic symmetries of the problem and realizability constraints. In these regards, apart from the Mix2 scheme, ML is the worst offender. The ML model is not constructed or trained to conserve ELN (see the complication described in Appendix~\ref{app:ELN_conservation}), and any ability to conserve ELN is learned from the training datasets themselves, which do conserve ELN to machine precision. Figure~\ref{fig:NSM_ELN_violation} shows CDFs of the violation of ELN conservation predicted by the ML model in each of the four unstable datasets. As in the previous results, the model struggles more with distributions that undergo more flavor transformation. In the NSM snapshots ELN is violated up to about 2.5 \%. 

\begin{figure*}
    \centering
    \includegraphics[width=\linewidth]{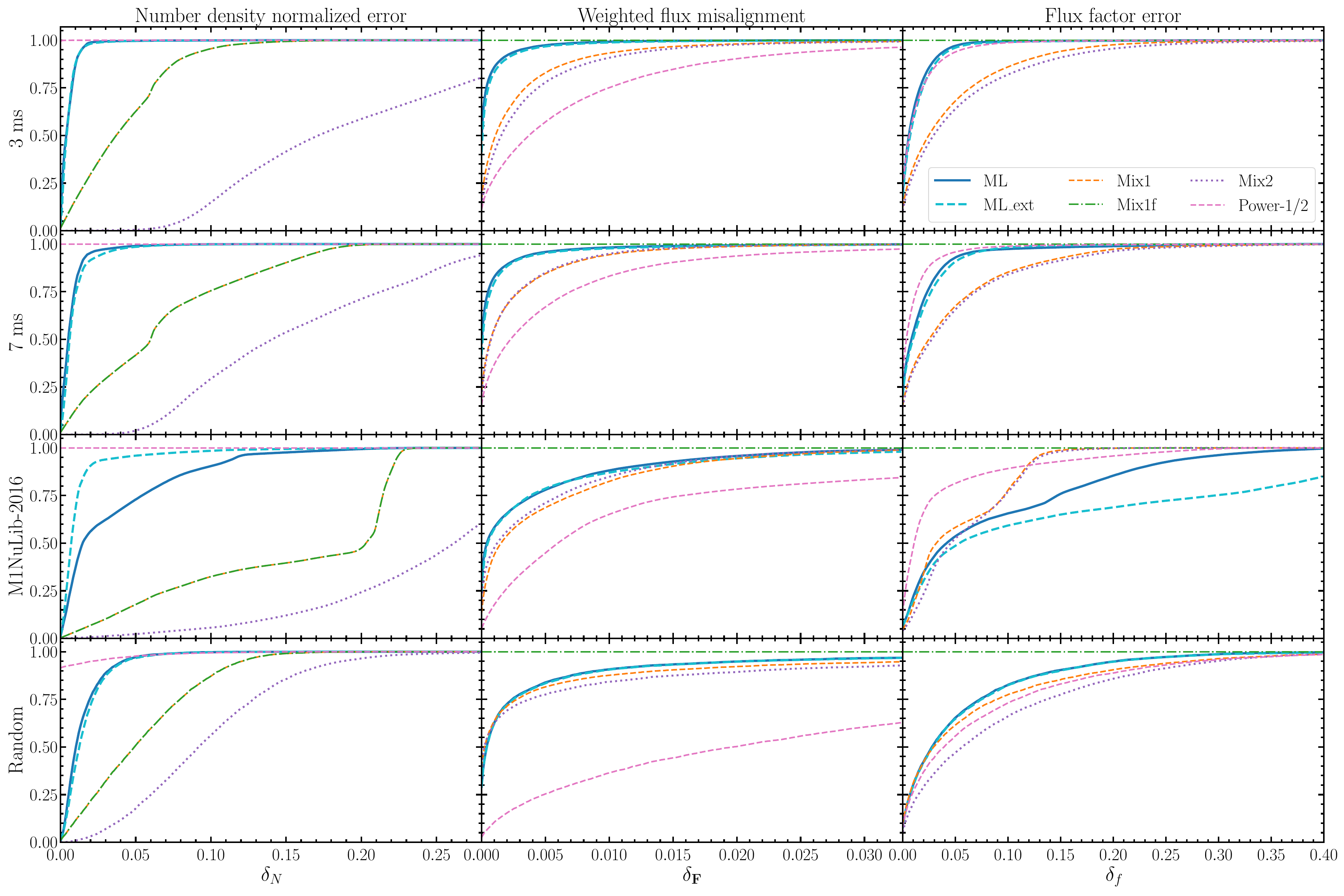}
    \caption{Cumulative distribution functions of the three error metrics on different datasets of \emph{stable} points, with the same plotting conventions as Fig.~\ref{fig:CDF_performance}. Here, no flavor transformation should take place, which is by construction satisfied by the Box3D model. The Mix1f scheme conserves flux factors and their directions, hence $\delta_{\vec{F}} = \delta_f = 0$ up to numerical precision. In this plot, the “stable” points for the 3 ms and 7 ms snapshots come from refinement levels 1,2,3 ; while the fiducial ML model was trained only on stable points from refinement level 1.}
    \label{fig:CDF_performance_stable}
\end{figure*}

The ML model very rarely makes unrealizable predictions. Namely, out of all the unstable datasets, no point leads to a negative number density, and 0.29 \% of the Random-unstable points lead to flux factors larger than one. The ML\_ext model has the same features, except that it predicts negative number densities in 0.5 \% of the M1NuLib-2016 unstable points probed, and flux factors larger than 1 in also 0.5 \% of these points (one third of these problematic points have both species with a negative number density and species with a flux factor larger than 1). If these issues do arise in a dynamical calculation, we suggest a minimal patch to the final prediction in Appendix~\ref{app:restrict_to_physical}.

\paragraph*{Performance on stable configurations —} In Fig.~\ref{fig:CDF_performance_stable} we plot distributions of prediction errors from applying the models to stable distributions. Ideally, the models should predict no flavor transformation in those cases. The Mix3D scheme is not plotted because, by construction, it predicts exactly no change when the distribution is stable. The ML model predicts number densities that change by at most a few percent on the training data (first, second, and fourth panels in the left column), but when the model is applied to unseen distributions (third panel in the left column) it performs much worse, with number density errors above 5 \% for a quarter of the data points. A similar trend is seen in predictions of the flux directions and flux factors. The ML\_ext model, which is also trained on stable points from the M1NuLib-2016 simulation, performs significantly better on this dataset as expected, with errors similar to those in the 3-ms and 7-ms datasets. Oddly, the ML\_ext model performs worse in predicting flux factors, despite being exposed to much of the M1NuLib-2016 dataset, as it seems to prioritize performance in predicting number density. Given the additional training data, the ML\_ext model may benefit from a larger neural network or from loss function tuning to prioritize performance appropriately.

The Mix1f scheme perfectly preserves the flux directions and flux factors, thus working as expected for stable distributions, but results in number density errors that are even larger than those predicted by the ML model. The Mix1 model predicts number densities identical to those predicted by Mix1f, but predicts fluxes that are much more wrong than those predicted by ML on average. Mix2 predictably performs very poorly in determining final number densities, since it predicts significant transformation regardless of the stability of the distributions. These issues are not a significant problem in the merger simulations where they were employed \cite{li_NeutrinoFastFlavor_2021,just_FastNeutrinoConversion_2022} because there was a binary flag determining whether a distribution is unstable, and flavor was only transformed if the distribution was deemed unstable. Thus, the ML model would benefit from using such a stability determination as well, and should produce results more accurate than the Mix models.

The Power-1/2 scheme correctly detects the stable points in the NSM datasets and hence predicts no change in number density. The errors in the flux metrics in the NSM datasets are then a result of removing the $x$ and $y$ components of the flux in the axisymmetrizing process. This results in flux alignment predictions that are generally worse, but flux factor predictions that are generally better than other schemes. However, it performs moderately worse in the (highly non-axisymmetric) random dataset, as it artificially predicts that some points are unstable in the axisymmetrized distributions (see bottom row, left panel of Fig.~\ref{fig:CDF_performance_stable}).

\paragraph*{Error topography ---} Several previous studies have indicated that the FFI occurs deep within the disk close to the energy-dependent neutrino decoupling surface. While the simplest models suggest ubiquity of the FFI \cite{Wu_FastNeutrinoConversions_2017}, investigation of the neutrino radiation fields in neutron star merger models based on initial accretion disk configurations suggest that there is quite ubiquitous instability at early times (i.e., tens of milliseconds) \cite{george2020fast}, but the FFI ceases in the polar regions with increasing time \cite{just_FastNeutrinoConversion_2022,Wu_ImprintsNeutrinoPair_2017}, though other models show the opposite trend \cite{li_NeutrinoFastFlavor_2021}. The complex trend of unstable regions following the distribution of tidal arms is only apparent in other calculations based on simulations starting from two distinct neutron stars instead of an initial equilibrium disk configuration \cite{Richers_evaluating_2022}. The largest errors in the ML model track these matter inhomogeneities, which suggests that fully three-dimensional, relativistic simulations of NSMs are a stronger testbed for flavor transformation subgrid models.

\begin{figure*}
    \centering
    \includegraphics[width=\linewidth]{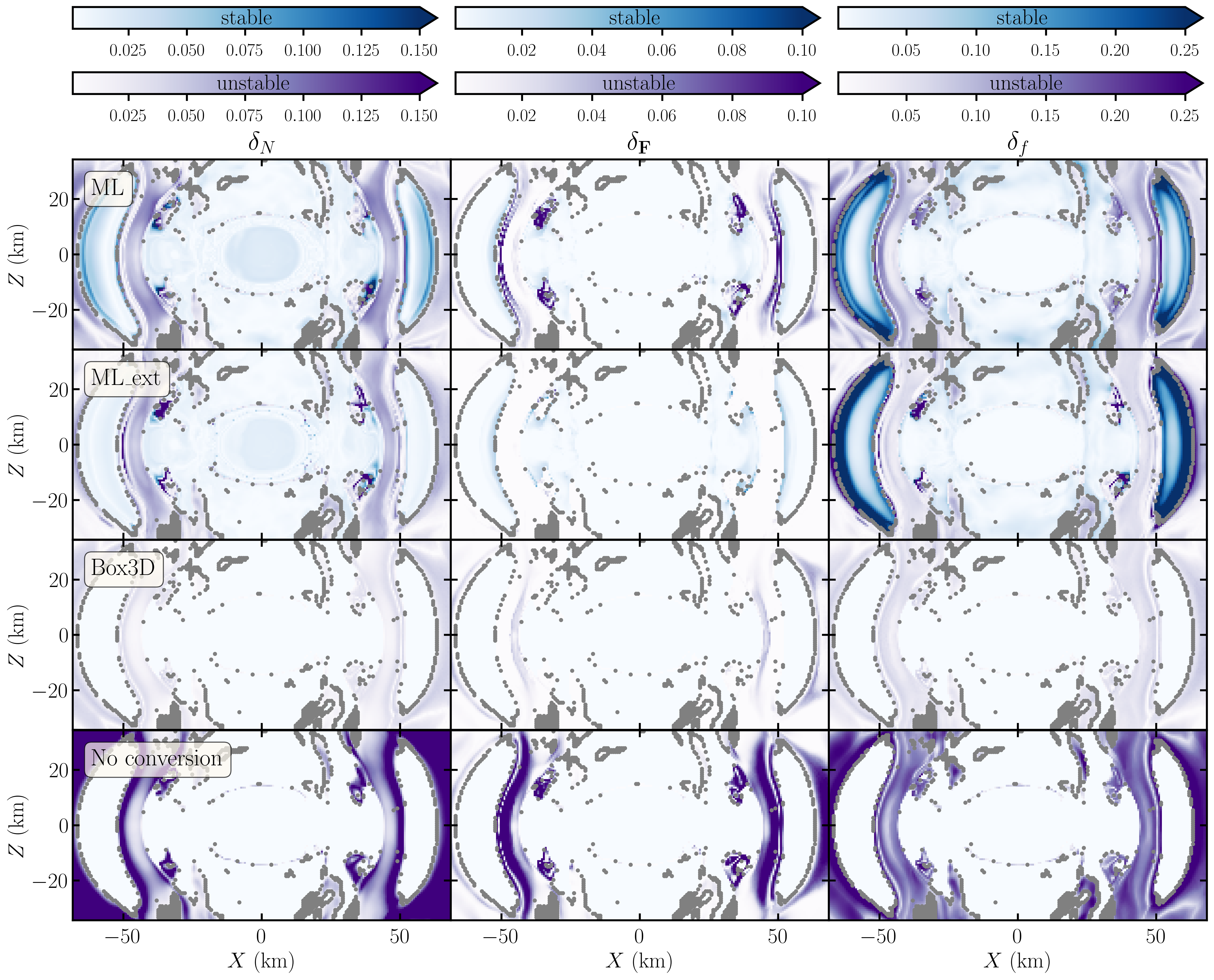}
    \caption{From top to bottom, errors from the application of the fiducial ML model, the extended ML\_ext model, the Box3D mixing scheme and if no flavor conversion is applied, visualized on the M1NuLib-2016 5 ms snapshot (see Fig.~\ref{fig:OldM1_snapshot}). The columns shows the error in Number Density ($\delta_N$, left), Flux misalignment ($\delta_\vec{F}$, middle), and the error in Flux factor magnitude ($\delta_f$, right). The error for stable points are represented in blue (the “true” final state being identical to the initial one) while the unstable points are shown in purple (the “true” final state being given by \texttt{Emu} simulations). The gray dots mark the unstable points for which the \texttt{Emu} simulations did not yield a result of sufficient quality.}
    \label{fig:performance_slices}
\end{figure*}

In this work, we only consider early-time snapshots due to the high cost of relativistic multidimensional NSM simulations. In order to quantify where the model predictions suffer most, we show the spatial distribution of errors in the M1NuLib-2016 snapshot induced by the ML and Box3D models in Fig.~\ref{fig:performance_slices}. We also include the errors incurred if no flavor conversion is considered (bottom row), which provides a measure of the actual amount of flavor transformation in the unstable points.

The blue color corresponds to the stable points while purple marks the unstable points with the “true” final value given by \texttt{Emu} simulations. The gray dots correspond to unstable configurations where the \texttt{Emu} simulations did not provide sufficient quality result, and largely populate regions of marginal instability. At 5 ms, the compact object at the center and the surrounding region out to about $30\,\mathrm{km}$  do not show the presence of FFI (with the exception of a small number of marginally unstable locations). The dense parts of the tidal arms in the equatorial regions at about $30$ and $50\,\mathrm{km}$ are also sources of neutrino emission and do not show instability, but in the space between the tidal arms there is a nontrivial interplay between radiation coming from the central compact object and from the nearby tidal arms. It is in these regions that the ML model has the most difficulty. As expected, the ML\_ext model performs better than the fiducial model on unstable points (consistently with Fig.~\ref{fig:CDF_performance}), and likewise in stable regions, with the exception of larger flux factor errors at the edge of instability (second row, right panel), once again in agreement with the CDFs in Fig.~\ref{fig:CDF_performance_stable}. As discussed before, the Box3D mixing scheme offers the best overall performance.

Recall that we apply the ML model to stable regions to thoroughly test its performance even though there should be no changed predicted in these distributions. The ML model predicts small changes in the central stable area, but incorrectly predicts large changes in the flux factors in the stable regions of the tidal arms. This area is the origin of the large errors seen in the right column of Fig.~\ref{fig:CDF_performance_stable}. Once again, the ML model would be improved by combining with a binary stability indicator as in \cite{li_NeutrinoFastFlavor_2021,just_FastNeutrinoConversion_2022}.

\section{Conclusion}
\label{sec:conclusion}

While simulations of core-collapse supernovae and neutron star mergers (NSMs) are expected to exhibit neutrino fast flavor instability, it is not yet understood how the instability modifies the dynamics and ejecta composition, since the requisite scales are impossible to resolve in global simulations. While pioneering calculations have bounded the magnitude of the FFI's impacts using a variety of subgrid models with qualitatively correct behavior, there are currently no subgrid models available that provide quantitatively correct predictions of how the FFI changes the densities and fluxes of each species of neutrino in a general multidimensional environment. In this work, we assess the accuracy of proposed subgrid models, extend a promising subgrid model to multiple dimensions, and create a machine learning (ML) model anchored in detailed small-scale simulations of the FFI.

To do so, we aggregated over 26,000 simulations of the FFI based on both randomized initial conditions and conditions extracted from global NSM simulations. In addition, we incorporated more than 286,000 neutrino distributions that are \textit{stable} to the FFI into the training and evaluation of our models.

The main objective of all of these models is to predict the first two angular moments (density and flux) of each neutrino species after saturation of the FFI using only the same angular moments as inputs. We constructed a ML model to do this based on a fully-connected neural network, but with additional structures to ensure conservation of particle number, to make the result rotationally invariant, and to make interpretation in the context of a fully general-relativistic NSM or supernova simulation straightforward. The code for constructing and training the ML model is open-source \cite{RheaCode}, as are the training data and the final trained models \cite{dataset}.

We also extend the box-like model of \cite{Zaizen:2022cik} to treat the FFI in non-axisymmetric neutrino distributions. Specifically, based on the pre-instability angular moments we construct a multidimensional distribution for each species, discretize each distribution into 10,000 directions, and apply a prescription to mix flavor at each direction based on the location of the ELN-XLN crossing.

The main result is that this “Box3D” model is particularly compelling as a subgrid model. It compares very favorably to the results of individual quantum kinetics simulations, both in terms of the final states of the angular moments of each species (red dashed line in Fig.~\ref{fig:NSM1}), and indeed the full angular distribution (bottom panel in Fig.~\ref{fig:final_distribs_NSM1}). Of the models studied, it most robustly results in small errors for predictions of the final states of unstable distributions (Fig.~\ref{fig:CDF_performance}), correctly predicts exactly zero change for stable distributions, and exactly conserves net lepton number. \footnote{Note that during the refereeing process of this paper, \cite{george2024_3D} presented an alternative generalization of the Power-1/2 scheme that appears more accurate than the Box3D scheme in their non-axisymmetric simulations.}

However, there are potential limitations to the Box3D model. First, it requires reconstructing and discretizing each distribution, and the resolution of the discretization can impact the results, especially if the distribution is highly forward-peaked. That is, even if the neutrino distribution is stable, the act of discretizing and re-integrating results in a change to the moments. This can be addressed in the case of stable distributions by only modifying neutrino distributions if they are unstable, but there are still resolution-dependent errors in unstable distributions. In addition, the model is specific to the FFI, and it is not clear if the model can be extended to more general situations outside of the approximations made in this work.

In this sense, the ML model is attractive, as it could a priori be trained on a wider range of flavor conversion mechanisms. It is, in addition, especially efficient on GPU-based systems. It exactly preserves net neutrino number, is rotationally invariant, and does not depend on a choice of discretization. Although all of the data we use to train the model conserves net lepton number, the model itself does not conserve lepton number, nor is additional loss imposed during training to drive conservation. Although the result is that the ML model incorrectly allows for violation of lepton number conservation in the FFI limit (Fig.~\ref{fig:NSM_ELN_violation}), it does approximately learn to conserve lepton number. In addition, various schemes can be employed to enforce lepton number conservation following the ML model's prediction (e.g., Appendix~\ref{app:ELN_conservation}). The ML model does, however, rarely make unrealizable predictions in the form of negative densities or flux factors larger than 1. This, too, can be accounted for after the fact with minimal modifications to the predictions (Appendix~\ref{app:restrict_to_physical}). The ML model is particularly worrying in that it predicts changes to both number densities and fluxes when applied to FFI-stable distributions (Figure~\ref{fig:CDF_performance_stable}). In real applications, it would benefit from a stability classifier (e.g., \cite{Abbar_SeachingFFI_2020,Nagakura_NewMethodDetecting_2021,Nagakura_WhereWhenWhy_2021,li_NeutrinoFastFlavor_2021,just_FastNeutrinoConversion_2022,Richers_evaluating_2022,Abbar_ML_detect_1,Abbar_ML_detect_2}) to enforce no change in stable distributions. 

For both the ML and Box3D models, errors seem to be largest for distributions that are marginally unstable (Fig.~\ref{fig:performance_slices}), while still being at the percent level for the Box3D scheme. This suggests the need for more dedicated studies of the outcome of the FFI for configurations at the edge of instability, as recently pointed out by~\cite{Fiorillo:2024qbl}. We leave this question for future work.

It is important to caution that good performance on training and test datasets drawn from the same statistical distribution (which is the case for the datasets taken from the NSM simulation in~\cite{Foucart:2024npn}) does not imply generalizability. This can be most clearly seen by comparing the results of our fiducial ML model to our ML\_ext model on the 5 ms M1NuLib-2016 NSM snapshot (third row of Figs.~\ref{fig:CDF_performance} and \ref{fig:CDF_performance_stable}). The fiducial ML model, which never sees these datasets, performs well below what one would predict based on its performance on test data during training. Not surprisingly, the ML\_ext model, which is also trained on data from this snapshot, shows much better performance. However, if applied to another completely independent NSM simulation, the ML and ML\_ext models would both deal with situations outside of the parameter space spanned by their training data, and could offer poorer performances. 
It should also be pointed out that FFIs can occur in other dense environments like core-collapse supernovae (see~\cite{Sen:2024fxa} and references therein). Notably, a number of the previous studies mentioned in this work were carried in spherically-symmetric CCSN environments (e.g.~\cite{bhattacharyya_FastFlavorDepolarization_2021,Ehring_FastNeutrinoFlavor_2023,Nagakura_BasicCharacteristics_2023,Xiong:2024tac}). It was nevertheless shown that non-axisymmetric ELN crossings can appear in CCSNe (e.g.,~\cite{Nagakura_FastCCSN_2019}), making the case for three-dimensional models like the Box3D prescription. While we expect this model should provide reasonable results in the context of neutrino distributions in supernovae, it is unclear how well the ML models will generalize from the NSM training sets to CCSN settings.

There are a number of further developments that could enhance the viability of subgrid models. First and foremost, the performance of the models needs to be compared with more global simulations of flavor instability to map out where the assumptions of instantaneous flavor change based on results from periodic box simulations is relevant, potentially requiring extensions to subgrid models to account for gradients. Even the models present here should be extended to account for other flavor transformation phenomena, although this will require a large number of additional calculations to use for training or assessment purposes. Although the Box3D model has a straightforward interpretation in the context of multi-energy distributions (the survival probability is the same for every neutrino energy along a given direction), the ML model would need to be verified or extended in this context. The simultaneous treatment of both lepton number and energy redistribution between neutrino flavors is not trivial, and care must be taken to treat both consistently. These are of course left for future work.

\begin{acknowledgments}

We thank Evan Grohs, James Kneller, Gail McLaughlin, David Radice, Ermal Rrapaj, and Don Willcox for useful discussions. S. R. was
supported by a National Science Foundation Astronomy and Astrophysics Postdoctoral Fellowship under Award No. AST-2001760. J.F. is supported by the Network for Neutrinos, Nuclear Astrophysics and Symmetries (N3AS), through the National Science Foundation Physics Frontier Center award No. PHY-2020275. S.G. is supported by the Nuclear Physics from Multi-Messenger Mergers (NP3M), through the National Science Foundation under Grant Number 21-16686. F.F. gratefully acknowledges support from the Department of Energy, Office of Science, Office of Nuclear Physics, under contract number
DE-AC02-05CH11231, from the NSF through grant AST-2107932, and from NASA through grant 80NSSC22K0719.

\end{acknowledgments}

\appendix

\section{Enforcing Additional Constraints in the ML Model}

In this appendix, we introduce various “fixes” that can be used to explicitly enforce some constraints or physical realizability conditions in the ML model.

\subsection{Conservation of Electron Lepton Number}
\label{app:ELN_conservation}

We could further modify the $\yML$ values in order to explicitly conserve net lepton number for each neutrino flavor. The initial and predicted net lepton number $L$ for each flavor $f$ is
\begin{equation}
    \begin{aligned}
        L^f &= \F^{t \nu f} - \F^{t\bar{\nu}f}
        &&= \left(\delta^\nu_m \delta^f_g - \delta^{\bar{\nu}}_m\delta^f_g\right)\F^{tmg}\\
        L^{f \prime} &= \F^{t \nu f \prime} - \F^{t\bar{\nu}f \prime} 
        &&= (\yML^{\nu f}_{mg} - \yML^{\bar{\nu} f}_{mg}) \F^{tmg}\,\,.
    \end{aligned}
\end{equation}
The excess of lepton number in the ML model prediction is then $L^{f \prime}-L^f$. If we subtract half of this excess from the neutrino distribution and add half to the antineutrino distribution, the model perfectly preserves lepton number. Specifically, we use a new transformation matrix given by
\begin{equation}
\label{eq:yhat_ELN}
    \begin{aligned}
        \widehat{\yML}^{\nu f}_{mg} &= \yML^{\nu f}_{mg} - \mathcal{E}^{f}_{mg}/2 \, , \\
        \widehat{\yML}^{\bar{\nu} f}_{mg} &= \yML^{\bar{\nu} f}_{mg} + \mathcal{E}^{f}_{mg}/2\,, \\
    \end{aligned}
\end{equation}
where
\begin{equation}
    \mathcal{E}^{f}_{mg} \equiv (\yML^{\nu f}_{mg}-\yML^{\bar{\nu} f}_{mg})-(\delta^\nu_m -\delta^{\bar{\nu}}_m) \delta^f_g \, .
\end{equation}
However, this causes all spacetime components of $\F$ to show this property, when in reality only the time component should be conserved. We refer to the model trained with this scheme as model ML\_yELN.

In addition, one can simply adjust the number densities directly to enforce ELN conservation. We refer to the model trained with this scheme as ML\_directELN. This is logically equivalent to imposing the above modification to the $\yML$ matrix only for the time component of the four-flux, as defined in the frame comoving with the background fluid, thus breaking the Lorentz invariance of our fiducial model. Although it may in fact be more correct to allow for a special reference frame, the use of this model is a bit more subtle in relativistic codes. In addition, care must be taken to transform energy density and number density consistently, since net lepton number is conserved, but the same is not true for neutrino energy. We explore this option as a curiosity here without proposing a generally consistent way to use it as a subgrid model.

As a proof of principle, we indicate in Table~\ref{tab:error_ELN} the error metrics of models trained with the same datasets as the fiducial ML model, but either using the $\widehat{y}$ matrix of Eq.~\eqref{eq:yhat_ELN} (“ML\_yELN” model), or with the direct modification of number densities (“ML\_directELN”). We only give the values for the M1NuLib-2016 snapshot, as they are representative of the overall trend. We introduced the notation $\delta_\mathrm{ELN} \equiv \left[(N'_{e,\mathrm{pred}}-\bN'_{e,\mathrm{pred}})-(N_e - \bN_e)\right]/N_\mathrm{tot}$, which corresponds to the quantity whose histogram is plotted on Fig.~\ref{fig:NSM_ELN_violation}.

\renewcommand{\arraystretch}{1.4}

\begin{table}[!ht]
    \begin{tabular}{|r|c|c|c|c|}
    \hline
    Model & $\mathrm{med}(\delta_N) $ & $\mathrm{med}(\delta_{\vec{F}})  $ & $\mathrm{med}(\delta_f)$ & $\mathrm{med}(\delta_\mathrm{ELN})$ \\
    \hline \hline
    ML & $0.0381$ & $0.0031$ & $0.0693$ & $0.0051$ \\ 
    ML\_yELN & $0.0259$ & $0.0058$ & $0.0957$ & $1.2 \times 10^{-8}$ \\
    ML\_directELN & $0.0207$ & $0.0031$ & $0.0604$ & $2.1\times 10^{-8}$ \\ \hline
    \end{tabular}
    \caption{\label{tab:error_ELN} Median (noted $\mathrm{med}$) of the error metrics and ELN non-conservation on the unstable dataset of the M1NuLib-2016 snapshot for the fiducial ML model and the two models adjusted for ELN conservation.}
\end{table}

First, as shown in the last column of Table~\ref{tab:error_ELN}, the “adjusted” models work as expected in that they assure ELN conservation. They offer, on average, better predictions for the output number densities, which is reasonable as ELN conservation \emph{is} satisfied by the true $N'$. Regarding fluxes, the ML\_directELN model performs similarly to the fiducial model, once again predictably since the fluxes are not affected by enforcing ELN conservation only in the number density moment. The change in the $\yML$ matrix leads, as explained above, to better predictions for the number densities, but worse ones for the fluxes, as they are incorrectly forced to parallel the change in number densities. This is particularly true in the “Random” dataset where axisymmetry is not satisfied and where fluxes change much more than in NSM environments (see Fig.~\ref{fig:CDF_performance}). Applying the ML\_yELN model to the Random dataset (not shown) results in $\delta_{\vec{F}}$ and $\delta_f$ errors one order of magnitude larger than the other models. It would therefore seem that directly changing the number densities to satisfy ELN conservation is the best strategy. However, this is not a viable path forward in general. First, we mentioned above the difficulties associated to relativistic invariance and the treatment of \emph{energy} densities on top of number densities. Then, ELN conservation is a special feature of fast flavor instabilities, such that a more general ML model must be able to not conserve ELN in some cases.

\subsection{Enforcing Realizability}
\label{app:restrict_to_physical}

In the event that the ML model makes an unphysical prediction (i.e., flux factor larger than 1 or negative number densities), it is possible to minimally modify the solution to bring it back to a realizable state. This is done by setting the final four fluxes to a weighted average between the complete mixing solution and the result of the ML model and choose the weighting such that $\mathcal{F}^{\alpha\prime\prime} \mathcal{F}''_\alpha\leq 0$ for all species (i.e., such that all four-fluxes are timelike). A flux factor larger than 1 would manifest as a spacelike four-flux, though this algorithm also works to address negative number densities as well. Specifically,
\begin{equation}
\label{eq:realizability_fix}
    \mathcal{F}^{\alpha n f \prime\prime} = \frac{\mathcal{F}^{\alpha n f \prime} + \sigma \langle\mathcal{F}'\rangle^{\alpha n}}{1+\sigma} \, ,
\end{equation}
where
\begin{equation}
\label{eq:realizability_sigma}
    \sigma \equiv \max_\text{species}
    \begin{cases}
    \dfrac{-b + \text{sgn}(a)\sqrt{r}}{2a} &  \text{if } |a/b|\geq 10^{-6} \, , \\
    -c/b & \text{if } |a/b|<10^{-6} \, ,
\end{cases}
\end{equation}
with $r=\max(b^2-4ac,0)$, $\langle\mathcal{F}\rangle^{\alpha n}=\sum_f\mathcal{F}^{\alpha n f}/3$ is the flavor-averaged four-flux, $a=\langle\mathcal{F}\rangle^{\alpha}\langle\mathcal{F}\rangle_\alpha$, $b=2\mathcal{F}^{\alpha}\langle\mathcal{F}\rangle_\alpha$, and $c=\mathcal{F}^{\alpha}\mathcal{F}_\alpha$, with primes and species indices suppressed. The piecewise nature of Eq.~\eqref{eq:realizability_sigma}, with the arbitrary threshold of $10^{-6}$, avoids floating-point precision errors near values of zero by neglecting the quadratic term when solving for $\sigma$. In addition, the fix is only applied if $\sigma>0$, in which case $10^{-6}$ is added to $\sigma$ before use in Eq.~\eqref{eq:realizability_fix}, which ensures the realizability of the final state in spite of potential round-off errors. We maximize over species such that we can use the same interpolation factor for all species to preserve net neutrino number conservation. The use of $\text{sgn}(a)$ selects the branch that leads to the most positive value of $\sigma$.

\bibliography{new_references}

\end{document}